\documentclass[12pt]{iopart}

\usepackage{iopams} 
\usepackage[utf8]{inputenc}
\usepackage{graphicx}
\usepackage{verbatim}
\usepackage{color}
\usepackage{url}
\usepackage{hyperref}

\begin{document}


\title[]{Standardization of type Ia supernovae}
\author{Rodrigo C V Coelho, Maur\'\i cio O Calv\~ao, Ribamar R R Reis and Beatriz B Siffert}

\address{Instituto de F\'\i sica, Universidade Federal do Rio de Janeiro,  Av. Athos da Silveira Ramos 149, 21941-972, Rio de Janeiro, RJ, Brazil}

\begin{abstract}
Type Ia supernovae (SNe Ia) have been intensively investigated due to its great homogeneity and high {luminosity}, which make it possible to use them as standardizable candles for the determination of cosmological parameters. In 2011, {the physics Nobel prize was awarded} ``for the discovery of the accelerating
expansion of the Universe through observations of distant supernovae.'' 
This a pedagogical article, aimed at those starting their study of that subject, in which we dwell on some topics related to the analysis of SNe Ia and their use in luminosity distance estimators. Here we investigate their spectral properties 
and light curve
standardization, paying careful attention to the fundamental quantities directly related to the SNe Ia observables. Finally, we describe our own step-by-step implementation of a classical light curve
fitter, the stretch, applying it to real data from the Carnegie Supernova Project.
\end{abstract}


\vspace{2pc}

\section{Introduction}
\label{sec:introduction}

A supernova (SN or SNe, from the plural \emph{supernovae}) is a stellar
explosion which may occur at the final stage of the evolution of a star or as the 
result of the interaction between stars in a binary system. The current 
supernova classification follows the historical order in which these events 
were observed. Initially, the explosions were divided into types I and II, 
according to the presence (type II) or absence (type I) of hydrogen emission 
lines in their spectra. Later, the observation of SNe with different spectral 
features resulted in the introduction of the subtypes which we use nowadays 
(see figure \ref{fig:SN_classification}). {SNe of types II, Ib and Ic are now 
believed to occur due to gravitational collapse of massive stars (above $\sim 
8$ solar masses), which leave behind a neutron star or 
a black hole. Type Ia SNe, on the other hand, are believed to be 
thermonuclear explosions in which the star is completely incinerated. {It 
is currently accepted that SN Ia are thermonuclear explosions \cite{Hoyle60} of carbon-oxygen 
white dwarfs that reach explosion conditions when, by accreting mass from a companion, 
approach the Chandrasekhar limiting mass {($\sim$ 1.4 solar masses)}\footnote{The exact value of 
this limiting mass depends on several properties of the white dwarf: 
metallicity, Coulomb corrections, temperature, rotation, magnetic fields, etc; 
in any event, realistically, these corrections seem to amount to no more than 
10\%.\cite{Padmanabhan01}}.\cite{Chandrasekhar31}
In the single degenerate scenario, the companion is generally considered to be a main sequence, 
a red giant or an AGB star, whereas in the double degenerate scenario it is another white dwarf. The nitty-grity details of 
the explosion process and the progenitor channel are 
still open to debate, both theoretically and 
observationally.\cite{Diemer13,Hillebrandt13,Maoz13}}

\begin{figure}[ht]
\center
\includegraphics[scale= 0.4]{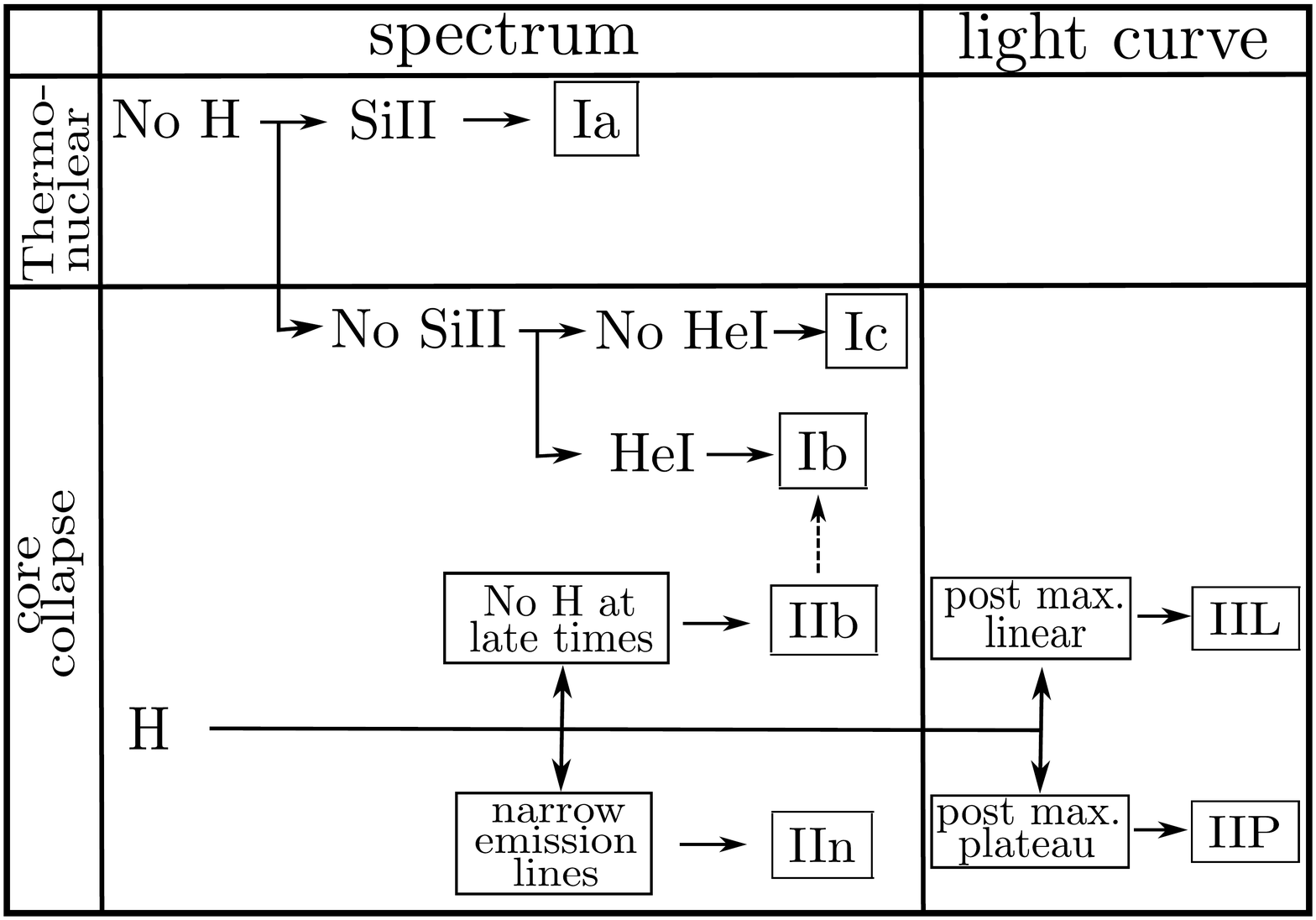}
\caption{Schematic classification of supernovae.}
\label{fig:SN_classification}
\end{figure}

As already stated, SNe are classified according to the presence or absence of 
certain spectral lines in their spectra and, for SNe II subtypes, the shape of their light curves. Type I SNe
can be divided into three subtypes: SNe Ia present a silicon absorption feature around
wavelength $\lambda=6150$ {\AA} (their main characteristic); SNe Ib does not
present silicon lines but present helium absorption lines; and SNe Ic present neither silicon nor helium
features. To learn more about the spectral features of SNe, see \cite{Filippenko97}. The most interesting subtype 
for cosmological purposes is the Ia, because of their high power, which
allows us to detect them in distant galaxies, and their quite homogeneous emission, which
makes possible their use as standard candles.  

A given class of astrophysical objects (or events) is considered a standard candle 
when their intrinsic luminosity is known or can somehow be estimated. In the case of SNe Ia, the 
observation of nearby events showed that all explosions had quite similar luminosities and the
relatively small {variations (as compared to the typical magnitudes of SNe Ia)} can be corrected 
for (in fact, due to the existence of such fluctuations these events should actually be considered \emph{standardizable} candles). SNe 
Ia themselves can be divided into subgroups and a classification scheme much 
used in the literature is the one by Branch \textit{et al.},\cite{Branch93}
according to which these events can be 1991bg-like\footnote{Supernovae are named for their year of occurrance and an uppercase letter, 
e.g., ``SN 1987A". If the alphabet is exhausted, double lower case naming is
used: [Year] aa .. az, ba .. bz, etc; e.g., ``SN 1997bs".}, which are subluminous, 
1991T-like, which are superluminous, and normal (\textit{Branch-normal}).
To have concrete numbers to express those {variations}, 
we calculated the sample standard deviation in absolute 
magnitudes $M_B$ (cf. section \ref{sec:fundamental_quantities}) of the SNe Ia in a sample of 
Vaughan \emph{et al.},\cite{Vaughan95} comprising 50 SNe Ia, of which 25 are Branch-normal. 
{Considering} only the 
Branch-normal SNe Ia, the standard deviation of the distribution 
of $M_B$ is 0.65 mag, while its mean is $-18.5$ mag}. For a more recent data sample, see \cite{Richardson14}
Since the flux of a source measured on Earth is proportional to the source's luminosity and
inversely proportional to its distance squared (more precisely, to its
luminosity distance squared, which we will define later), we see that it is
possible to estimate the SNe Ia distance by measuring its flux.  

{It was using SNe Ia that the winners of the 2011 physics Nobel prize 
discovered in 1998 and 1999 that
the universe is currently on an accelerated 
expansion.\cite{Riess98,Perlmutter99}} 
Currently, we believe that the cosmic acceleration is caused, in the context of general relativity,
by an unknown form of energy, called dark energy, which would generate a gravitational repulsion
unlike radiation, baryonic and cold dark matter, for which the gravitational interaction is
attractive. The most popular candidate for dark energy is the cosmological constant, which is usually 
interpreted as vacuum energy. There are, nevertheless, other explanations for the accelerated expansion being 
investigated, based either on modifications of general relativity, or on the existence of inhomogeneities 
in the matter distribution of the Universe.\cite{Kunz12}

As already mentioned, SNe Ia are standardizable candles which can be observed in very distant galaxies
due to their high power. They are, however, rare events \footnote{For supernovae relatively close to our galaxy
with $0<z<0.3$, the rate of occurrence of SNe Ia per volume is 
$(3.43\pm0,15)\times 10^{-5} \mbox{
supernovae/year/Mpc}^{3}$, according to \cite{Dilday10}.} and, since being explosions, they
are transients (lasting around three months), which makes their observation a difficult task. In order
to detect a high number of SNe, various projects are being planned, as this will demand a greater 
number of researchers in the field. For a list of these projects and some of the most important past and 
present experiments, see Table \ref{tab:SNexp}. Our goal in this work is to highlight some basic concepts
concerning the use of SNe Ia for cosmology, which we found are not detailed in textbooks. 
We believe that this work will be of great utility for those who are starting their research 
in the field, as well as for researchers who have never worked specifically in this field.

\begin{table}
\centering
\caption{Past, current and future experiments to detect SNe: Equation of State: SupErNovae trace Cosmic Expansion (ESSENCE) \cite{Miknaitis07}, Supernova Legacy Survey (SNLS) \cite{Conley11}, Sloan Digital Sky Survey (SDSS) \cite{Sako14}, Panoramic Survey Telescope \& Rapid Response System (Pan-STARRS) \cite{Rest13}, Dark Energy Survey (DES) \cite{Bernstein12}, Javalambre Physics of the Accelerating Universe (J-PAS) \cite{Benitez14} and Large Synoptic Survey Telescope (LSST) \cite{Abell09}. The third column gives the number of spectroscopically confirmed SNe Ia for past experiments and the total number of expected detections for current and future ones.}
\footnotesize
 \begin{tabular*}{\textwidth}{@{}l*{15}{@{\extracolsep{0pt plus12pt}}l}}
 \br
 Name & Running period & \# of SNe Ia & Redshift range\\
 \mr
 ESSENCE & 2002--2007 & 102 & 0.1--0.78\\
 SNLS & 2003--2008 & 242 & 0.3--1.0\\
 SDSS & 2005--2007 & 448 & 0.05--0.4\\
 Pan-STARRS & 2009--2014 & $\sim$ 3300 & 0.03--0.65\\
 DES & 2013--2018 & $\sim$ 4000 & 0.05--1.2 \\
 J-PAS & 2015--2021 &  $\sim$ 3800 & 0.05--0.4\\
 LSST & 2022--2032 & $\sim$ 10000 per year & 0.1--1.2\\
 \br
 \end{tabular*}\\
 \label{tab:SNexp}
\end{table}
\normalsize

{In section~\ref{sec:fundamental_quantities} we present the basic concepts of
spectrum, light curve, flux, magnitudes, all of them derivable from the
fundamental concept of specific flux. In section~\ref{sec:dependence}, we discuss
the influences distance and redshift have on the specific flux of an arbitrary
source. In section~\ref{sec:LC_standardization}
we discuss our naive light curve standardization, by taking advantage of a sort of
stretch correction that characterizes only the variations in SNe Ia rise-and-decline rates, but not the
intrinsic luminosity differences. In section~\ref{sec:conclusion} we present our
conclusions. In \ref{sec:basic_transformations} we describe
some usual transformations  of an
arbitrary function, for generic pedagogical
reasons.}

\section{Fundamental quantities} 
\label{sec:fundamental_quantities}

The specific flux \footnote{The expression \emph{specific} refers to quantities
measured per unit wavelength (or frequency), while \emph{bolometric} refers to
quantities integrated over all wavelengths (or frequencies).} (in the wavelength representation) measured by a detector is generically defined as the infinitesimal energy received by
the detector per infinitesimal time interval, per infinitesimal perpendicular
area, per infinitesimal wavelength interval,\footnote{The typical unit of $f_\lambda$ is
1 erg/cm$^2$/s/\AA, whereas for the corresponding frequency
representation, $f_\nu(t,\nu)=cf_\lambda(t,c/\nu)/\nu^2$, it is 1
erg/cm$^2$/s/Hz = $10^{23}$ Jy (jansky).} i.e.,
\begin{equation}
 f_\lambda := \frac{dE}{dt\,dA_\perp\,d\lambda}\,.
\label{specific_flux_definition}
\end{equation}

The specific flux for a given source will in general depend not only on the wavelength $\lambda$, on the distance to the source $r$ (cf. subsection ~\ref{subsec:distance}) and on the source's specific power or luminosity $L_{\lambda}$, but also, for transient sources, on the time $t$, and for moving sources, on the redshift $z$ (cf. subsection ~\ref{subsec:redshift}); concretely $f_{\lambda}=f_{\lambda}(\lambda, t, r, z, L_{\lambda})$. For
simplicity, in future references to this equation we may suppress one or more dependences
in the function $f_{\lambda}$. More on the discussion present in this section can be found in classical astrophysics books such as \cite{Carroll06}.

\begin{figure}[ht]
\center
\includegraphics[scale=0.35]{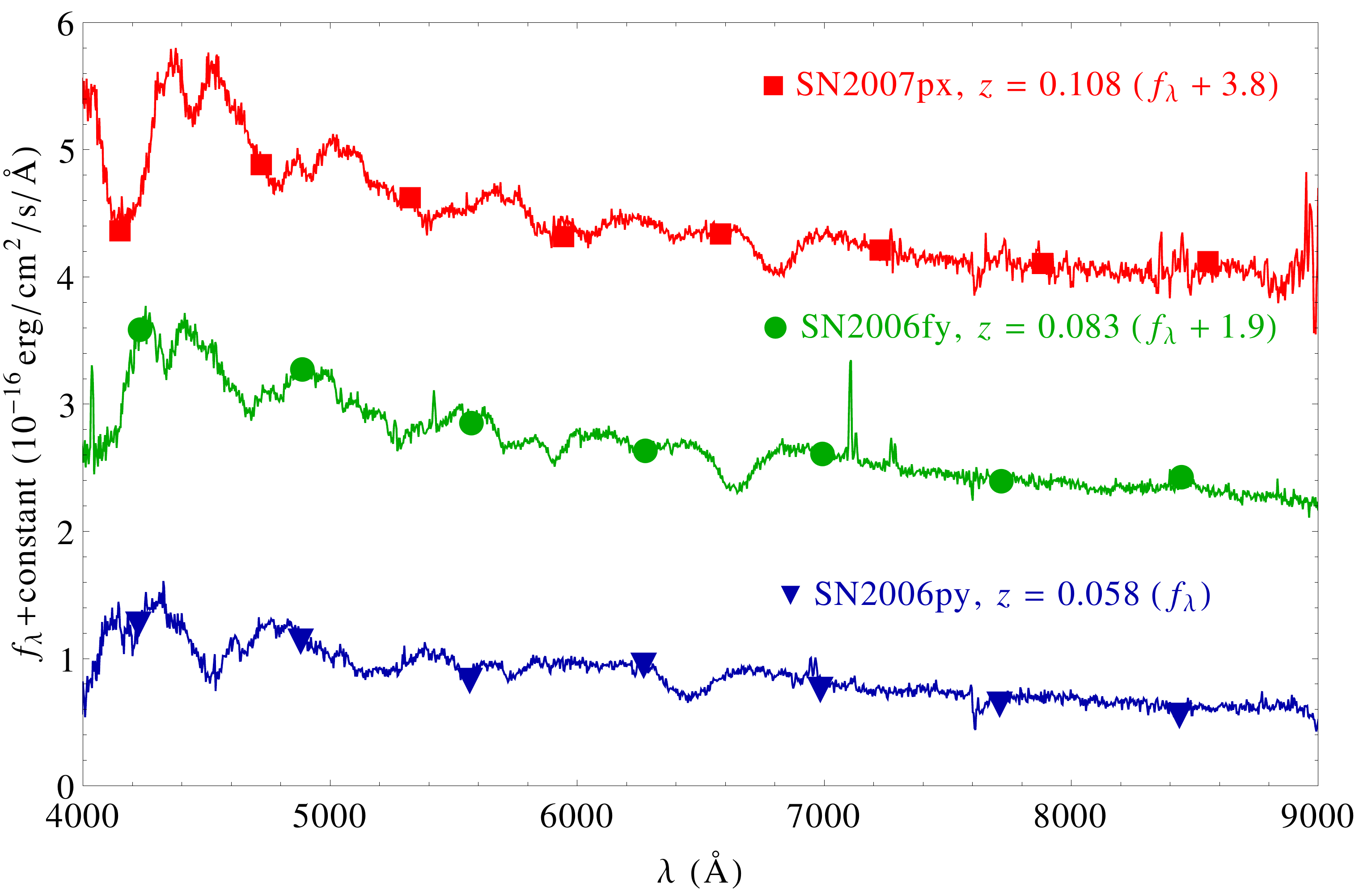}
\caption{Observed spectra of some SNe Ia, at four
days before $B$ band maximum light.\cite{Ostman11} Here and in all forthcoming 
figures showing spectra, geometric symbols {(circles, squares, and triangles)} serve only as a guide for the reader to
better identify to which SN each curve refers.}
\label{fig:typical_spectra}
\end{figure}

{There are basically two techniques used for detecting astronomical objects:
spectroscopy and photometry. In spectroscopy one uses a spectrograph to
decompose the incoming light into its different wavelength components and obtain
a measure of the specific flux at a given time, i.e. the spectrum of the object.
Despite the {high spectral resolving power in wavelength
($R:=\lambda/\Delta\lambda$, where $\Delta\lambda$ is the resolution of the spectrograph) provided by spectroscopy (a low to intermediate 
resolution spectrograph has $R$ of the order 1000--10000, whereas state of the
art high resolution ones can achieve $R\simeq 100000$)}, it demands more
observation time per object and more expensive equipment. In
figure~\ref{fig:typical_spectra}, we show some spectra from typical SNe Ia.}

In photometry one uses filters, which let the light pass only for a particular
wavelength
interval (the filter bandpass), and the resulting observation, called flux, 
corresponds to
specific flux 
integrated over this interval. Flux measures in a given filter at different
times (or epochs)
constitute a usual (not specific) light curve of the object.
Photometry is a cheaper and faster technique and there are many projects being
designed to get a huge amount of data through photometric observations.

We now show how to obtain SN Ia light curve templates at a given
filter from a theoretical model for $f_{\lambda}(\lambda,t)$. These templates are
necessary for the standardization of SN Ia light curves, as will be discussed
in section \ref{sec:LC_standardization}.

\begin{figure}[ht]
\center
\includegraphics[scale= 0.25]{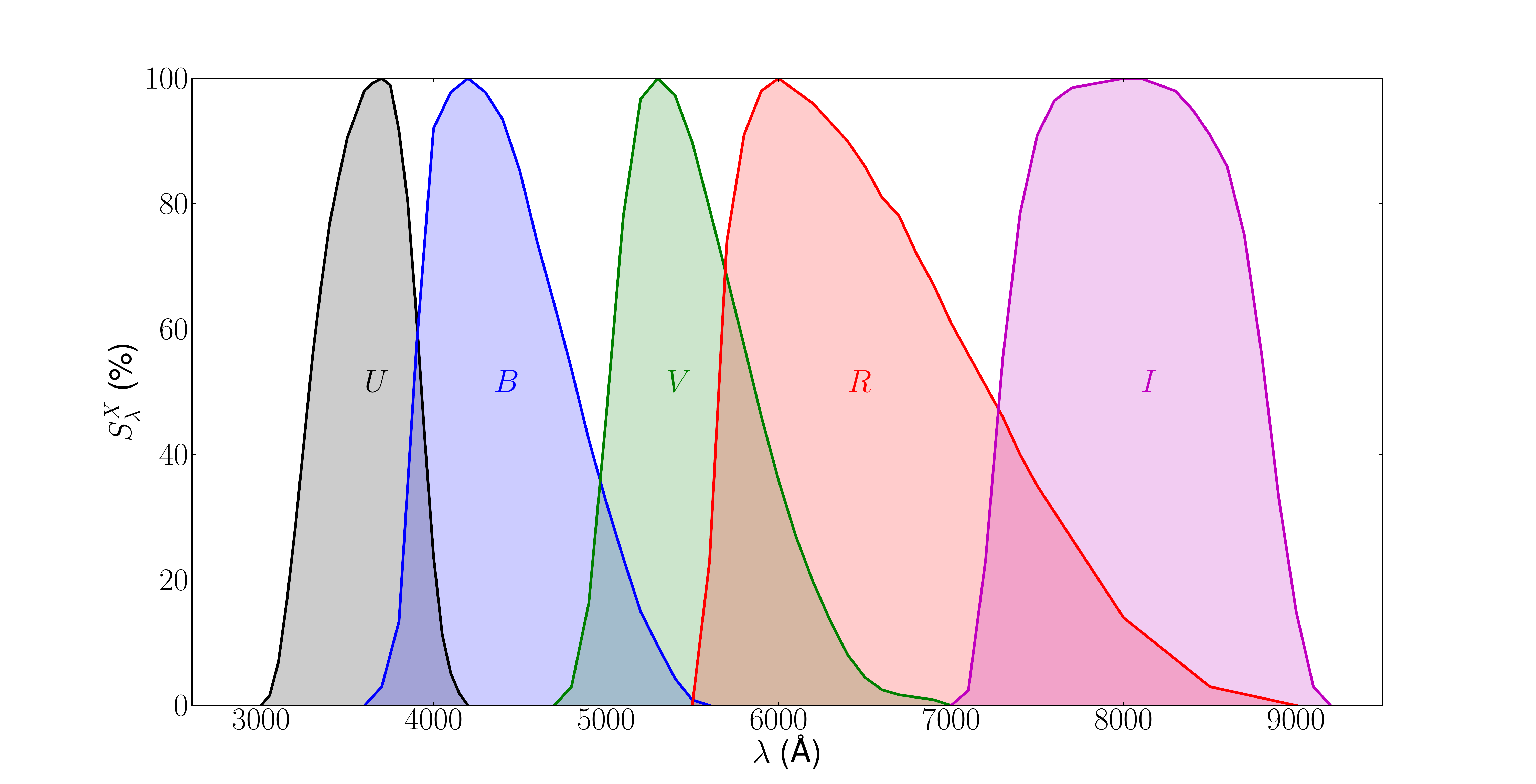}
\caption{Transmissivity curves for the \textit{UBVRI} filters typically used in
photometry.}
\label{fig:filters}
\end{figure}

First we have to take into account the bandpass of the chosen
filter. The filters \textit{UBVRI} (also known as the Johnson-Cousins filter set) are traditionally used to characterize SNe in
the rest frame and will be used in this work. The reader can find a detailed discussion on photometric systems in Bessell.\cite{Bessell05} We show in figure \ref{fig:filters}
the transmissivity curves, i.e. the fraction of energy that passes through
the filter as a function of wavelength, $S^X_\lambda$, for these filters. It is
important to notice that the filters are not perfect, in the sense that they do
not let all photons pass, no matter what wavelength we consider. {We will
define the flux in band $X$, $f^X$, as the energy flux that is transmitted
 through filter $X$, which can be written as}
\begin{equation}
f^X(t) := \int^\infty_0 f_\lambda(\lambda, t) S^X_\lambda(\lambda) d\lambda,
\label{fband}
\end{equation}
where we have, for brevity of notation, suppressed the dependence of 
$f_\lambda$ (and thereby of $f^X$) on $\boldsymbol r$.

The light curves are generally given in terms of the apparent magnitude in a
given filter $X$, which is related to the flux $f^X$ by\footnote{Throughout the 
text $\log$ denotes decimal (base 10) logarithm.}
\begin{equation}
 m_X(t) := -2.5 \log \left (\frac{ f^X(t) }{g^X} \right ),
 \label{mx}
\end{equation}
where $g^X$ is the reference flux, that can be for instance the flux of a given
star to which 
all other sources will be compared and defines a \emph{magnitude system}. A
\emph{photometric system} is defined by a set of filters (in our case
\textit{UBVRI}) and
the reference flux defined in all of them. In principle, the reference flux
can be different for each filter, however this is not mandatory. In this work we
use the \textit{AB} magnitude system,\cite{Oke65,Oke83} which uses as reference
a
constant specific flux
for all frequencies:
$$g^{AB}_\nu = 3631 \mbox{ Jy}\,.$$
Another commonly used magnitude system is the one that uses as reference flux
the flux of the Vega star in the chosen filters. Our photometric system will be
defined by the filter set \textit{UBVRI} and the \textit{AB} magnitudes. In 
order to mantain the notation most commonly used by astronomers, throughout the 
text we are going to refer to the apparent magnitude in a given filter $X$ by simply 
the letter $X$ so, for example, the apparent magnitude of an object measured with the 
$B$ filter will be just denoted $B$.

The filter reference flux $g^X$ is given by
\begin{equation}
g^X := \int^\infty_0 g_\lambda^X(\lambda) S^X_\lambda(\lambda) d\lambda,
\label{gbanda}
\end{equation}
where $g_\lambda^X(\lambda)$ is the specfic reference flux for filter $X$.

Since we chose to perform our calculations in wavelength space, we need to
rewrite the \textit{AB} reference specific flux using the relation
$$g_\nu(\nu) d\nu = g_\lambda(\lambda) d\lambda.$$
Recalling that $c = \lambda \nu$, we can obtain the reference specific flux as 
a function of wavelength
$$g_\lambda^X(\lambda)= \frac{cg^{AB}_\nu}{\lambda^2}.$$

\begin{figure}[ht]
\center
\includegraphics[scale=0.35]{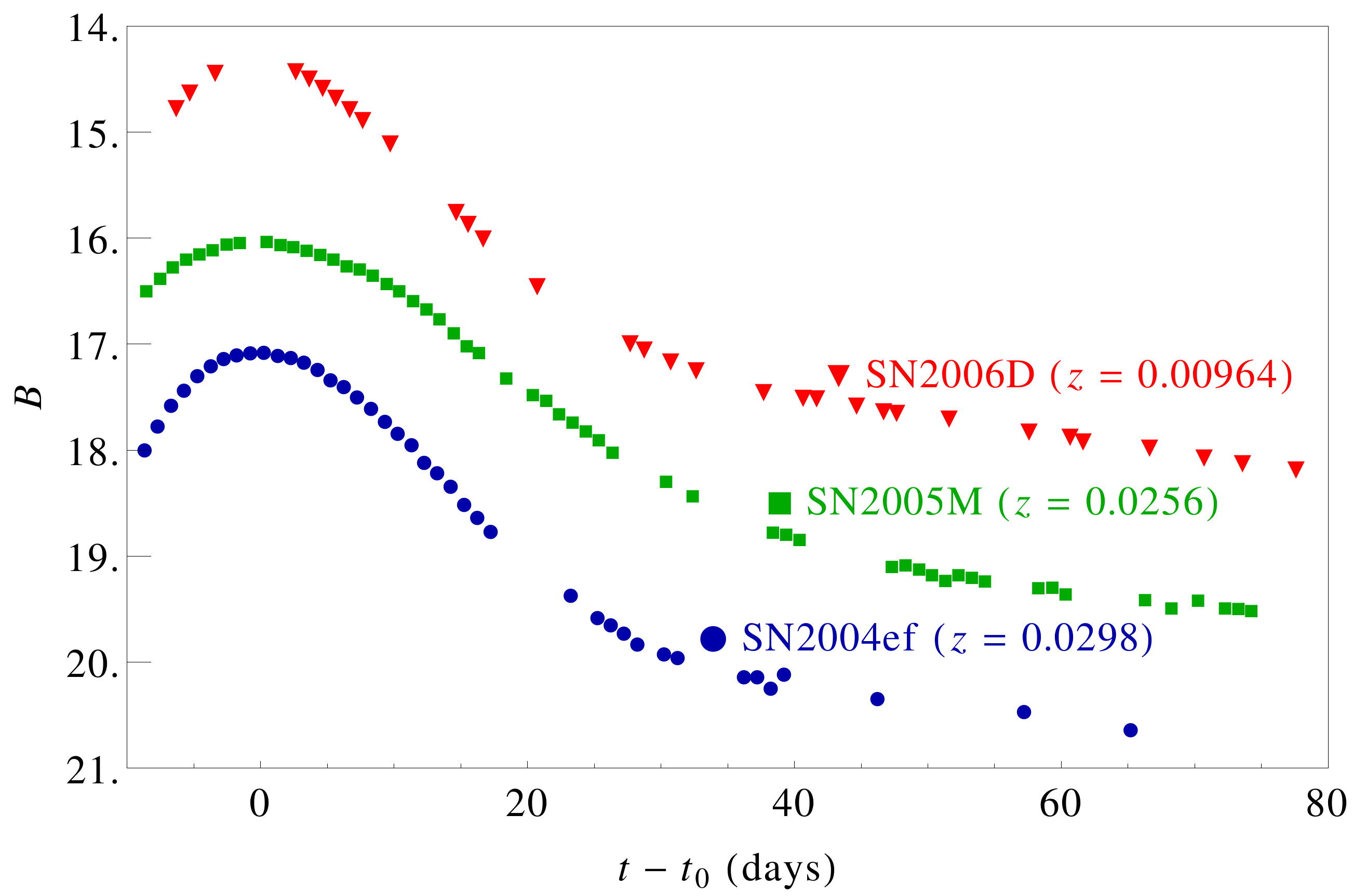}
\caption{{Observed sampling of apparent magnitude
$B$ band light curves from two Branch-normal (SN2004ef and SN2006D) and one
1991T-like (SN2005M) SNe Ia.\cite{Contreras10} Notice that a simple visual
inspection of the light curves does not allow determining the subtypes.}}
\label{fig:typical_light_curves}
\end{figure}

\begin{figure}[ht]
\center
\includegraphics[scale= 0.6]{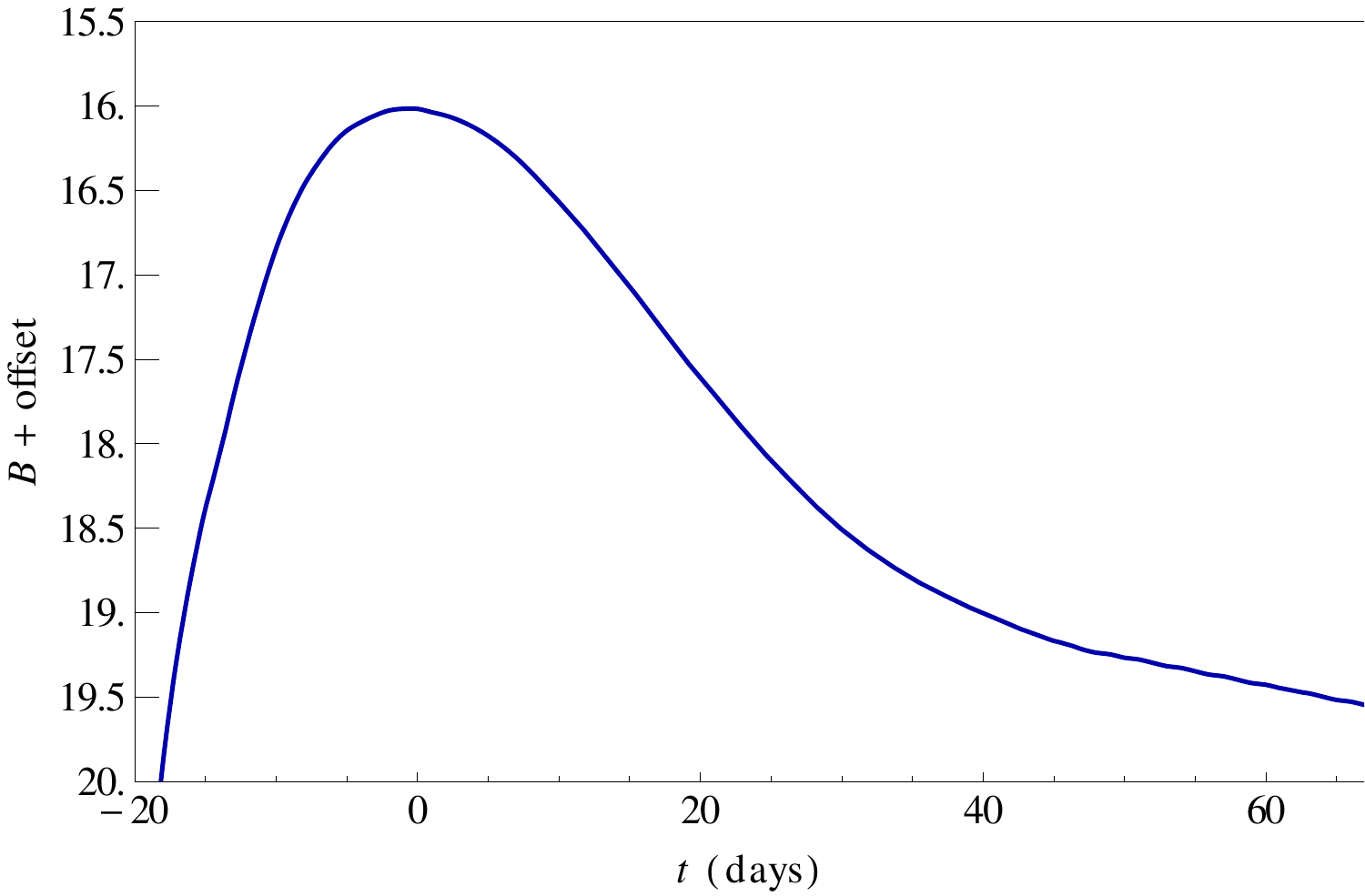}
\caption{{Theoretical $B$ band light curve
constructed from Nugent's Branch-normal SN Ia spectral 
template.\cite{Nugent_templates}}
}
\label{fig:template}
\end{figure}

\noindent{Therefore, to build a light curve, we need to evaluate the magnitudes
for a
given filter using (\ref{mx})} for spectra at different epochs. In
figure~\ref{fig:typical_light_curves}, we show some light curves from typical SNe
Ia, whereas in figure \ref{fig:template} we show a SN Ia light curve obtained
from the SN Ia template generated by Nugent.\cite{Nugent_templates}

\begin{figure}[ht]
\flushleft
\includegraphics[scale= 0.32]{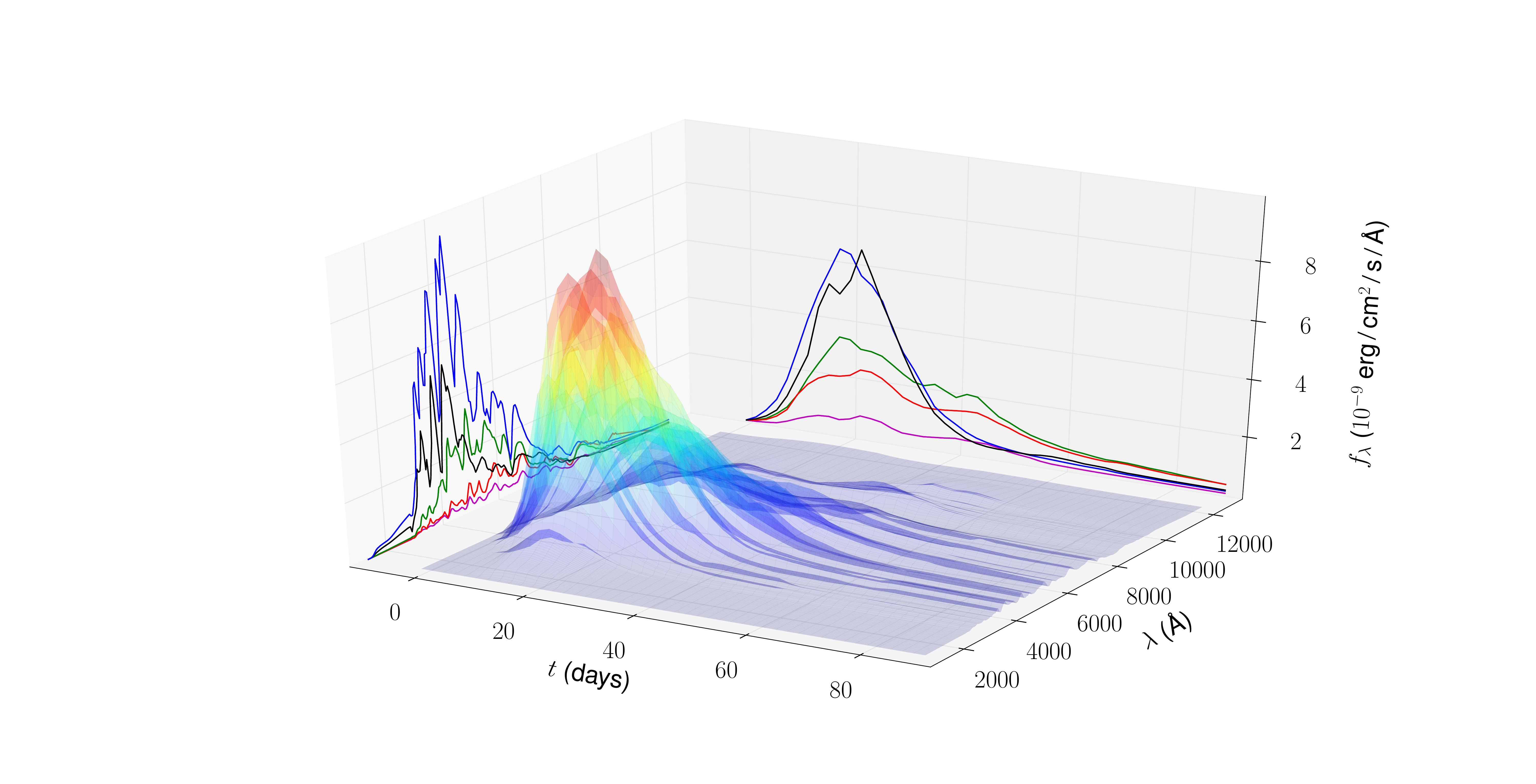}
\caption{Theoretical rest-frame spectral surface generated from Nugent's
template of synthetic spectra, at different epochs or phases, of a typical 
Branch-normal SN Ia.\cite{Nugent_templates} We also show five
typical spectra, projected onto a conveniently offset plane $t=-10$ days, and
five typical specific light curves, projected onto another conveniently offset
plane $\lambda=13000$ \AA.}
\label{fig:sedsurface}
\end{figure}

With given source and detector, we can display the
visual representation of the function $f_\lambda(\lambda, t)$ by means of what
we will call the spectral surface. In figure~\ref{fig:sedsurface}, for instance,
we show a representation of this surface, constructed from Nugent's estimates based on real SNe Ia data,\cite{Nugent_templates} for fixed $z,r$ and $L_{\lambda}$ (cf. section \ref{sec:dependence}).
The spectral surface displays in one single frame both the time evolution of the spectrum and the wavelength dependence of the specific light curve. The spectrum of the source, at a given
time $t_*$, is the intersection of the spectral surface with the plane
$t=t_*$, and the specific light curve, at a given wavelength $\lambda_*$,
 is the intersection of the spectral surface with the plane
$\lambda=\lambda_*$. 
A spectral surface like the one shown in figure \ref{fig:sedsurface} would be the
result of ideal observations of a SN, continuous both in wavelength and time.
In practice the best we can do is a discrete sampling of that surface
for a given SN; however, even this would be unfeasible for a high number of
SNe, because of the time demanded for the observations and the need for high
cost facilities. 

{It is convenient to find a relationship between ideal detected quantities
and intrinsic (source rest-frame) ones in a cosmological spacetime. To that end,
as a motivating warm-up, let us consider an imaginary spherical
(2-dimensional) surface, of radius $R$, concentric with a light source, both at 
rest in an
inertial frame of the Minkowski spacetime. The
bolometric (raw or pure) flux is defined 
as
\begin{equation}
f(t,R,L_{\lambda}) := \int_{0}^\infty
f_\lambda(\lambda,t,R,L_{\lambda})\,d\lambda\,, 
\label{fbolometric}
\end{equation}
and, due to conservation of energy, is trivially related to the intrinsic
bolometric luminosity $L(t):=\int_{0}^\infty
L_\lambda(\lambda,t)\,d\lambda$\footnote{Bolometric flux has the same units as band-limited
flux: 1 erg/cm$^2$/s.} by:
\begin{equation}
f(t, R, L) = \frac{L(t)}{4 \pi R^2}\,.
\label{bolometric}
\end{equation}
} 

We now introduce the concept of the redshift $z$, which is a measure 
of the relative velocity between astrophysical objects through 
the observation of their spectral features \cite{Synge60,Narlikar94}:

$$z:=(\lambda_{obs}-\lambda_{em})/\lambda_{em},$$
where $\lambda_{em}$ is the wavelength of a spectral feature, as measured in its rest frame, and
$\lambda_{obs}$ is the corresponding wavelength measured on Earth.

In \ref{sec:basic_equation} we show an intuitive way to obtain 
the relation between flux and luminosity for a more general spacetime, taking $z$ into account,
which is (\ref{specific_flux})

\begin{equation}
f_\lambda(\lambda, t, r, z, L_\lambda) = \frac{L_\lambda \left (
\,{\lambda}/{(1 + z)}, {t}/{(1+z)}\, \right)}{(1 + z)^3 4 \pi r^2},
\end{equation}
where $L_\lambda(\,{\lambda}/{(1+z)},{t}/{(1+z)}\,)$ is the specific 
luminosity in
the source's rest-frame.  

\begin{figure}[ht]
\center
\includegraphics[width= 0.46\textwidth]{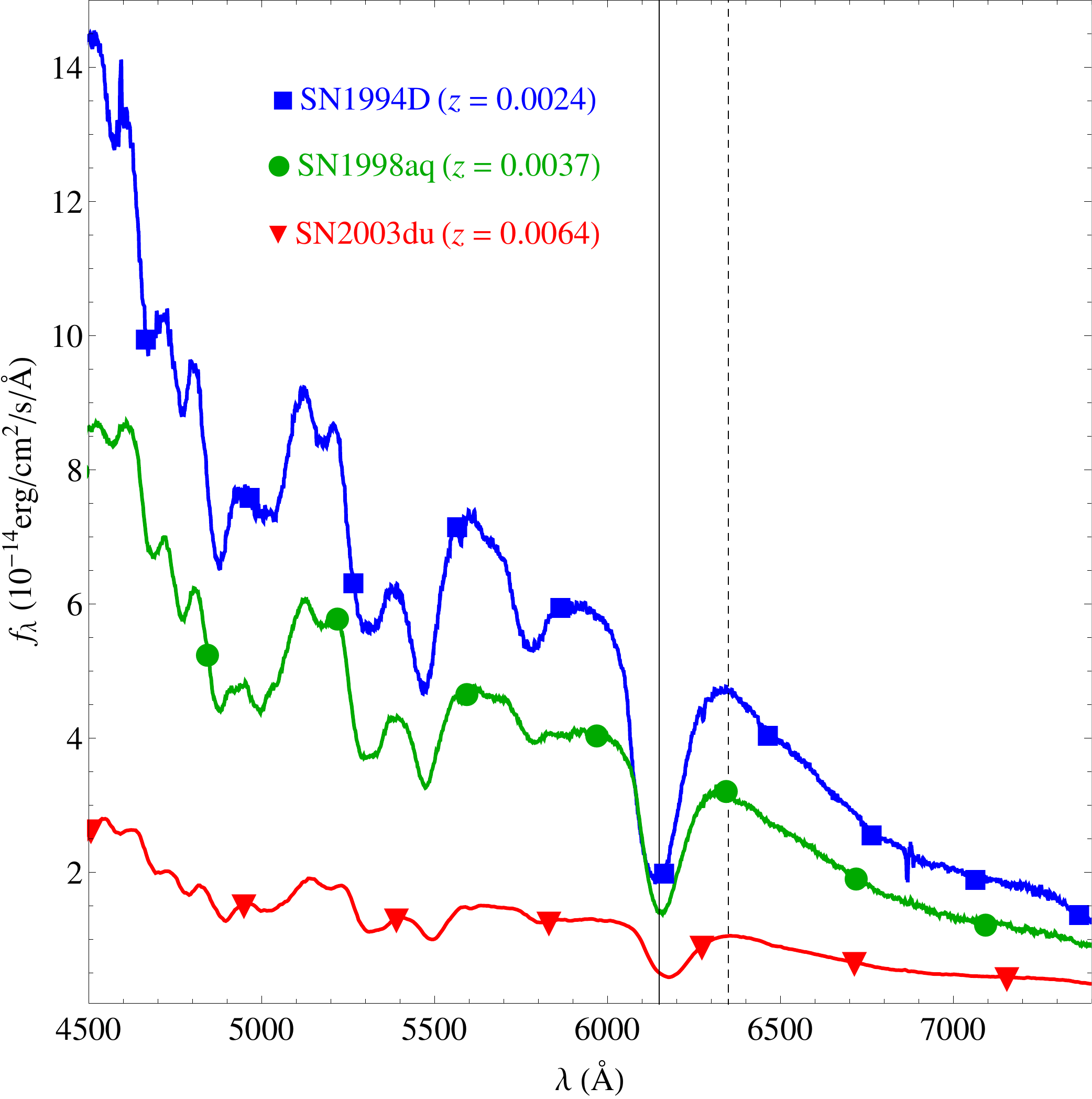}
\qquad\includegraphics[width = 0.46\textwidth]{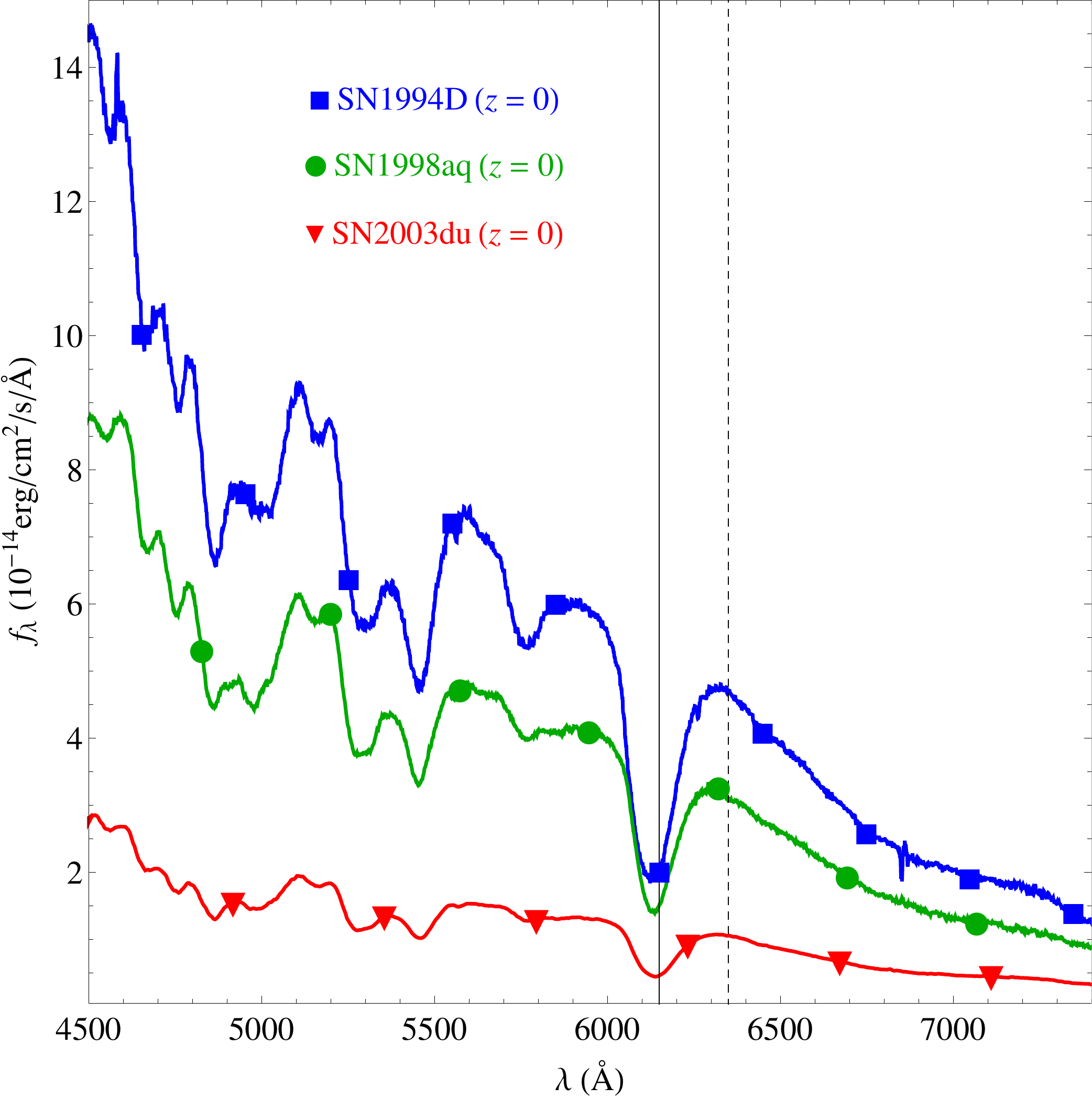}
\caption{{\textit{Left panel}: spectra from 
Branch-normal SNe Ia 1994D, 1998aq and 2003du taken two days
after maximum, in the $B$ band. \textit{Right panel}: same spectra in the SN Ia
rest frame (cf. subsection~\ref{subsec:redshift}). 
The vertical solid (dashed) lines indicate the typical rest-frame position of
the absorption (emission) components for SiII, due to the P Cygni profile.}}
\label{fig:d}
\end{figure}

From specific flux measures in different wavelengths (or frequencies), in a given epoch, 
we can construct a spectrum of an astrophysical object. In figure \ref{fig:d},
left panel, we
show spectra of three SNe Ia, SN1994D,\cite{Patat96} SN1998aq\cite{Branch03}
and 
SN2003du,\cite{Gerardy05} taken two days after maximum light (in $B$ band,
as we will see in the next sections), from the public database SUSPECT.\cite{suspect}
{The characteristic shape of the spectral lines, known as P Cygni profile, 
indicates
the presence of an expanding gas cloud. For a {gas expanding with spherical symmetry}, part of the light
that is emitted toward us is coming from regions that are moving in our direction and is 
\emph{blueshifted}, and the other part comes from regions that are moving away
from us{, being therefore redshifted. Since different layers of the
expanding gas move with different velocities, the resulting spectrum presents
wide emission lines centered at the rest wavelenghth value.} As an example of 
such
lines, we can see two SiII absorption lines with rest-frame wavelengths $\lambda
\approx 6347$ {\AA}
and $\lambda \approx 6371$ {\AA} which appear in the spectrum in figure
\ref{fig:d} as a
broad absorption feature at $\lambda \approx 6150$ {\AA} (indicated by the
dashed
vertical line) followed by an emission line centered at $\lambda \approx 6350$
{\AA}.}

Based on the definition of apparent magnitude in a given filter, given by (\ref{mx}),
we can introduce the concept of absolute
magnitude, the apparent magnitude that the source would have for a
hypothetical observer at a distance of $10$ parsecs\footnote{The parsec is a distance unit frequently used in astronomy
and corresponds to approximately $3.26$ light-years or $3.08 \times10^{16}$ m.
1 parsec is the distance to an object with rest-frame size of 1 astronomical unit
and apparent angular size of 1 arc second.} and at rest
with respect to it ($z=0$),

\begin{equation}
M_X(t) := -2.5 \log \left (\frac{ \displaystyle\int^\infty_0
\frac{L_\lambda(\lambda,t)}{4 \pi (10 {\rm\ pc})^2}
S^X_\lambda(\lambda) d\lambda }{\displaystyle\int^\infty_{\lambda=0}
g^{X}_\lambda(\lambda)
S^X_\lambda(\lambda) d\lambda} \right ). 
\label{me}
\end{equation}
{We would like to call the reader's attention to the fact that, by its very definition, it makes no sense to refer to an absolute magnitude for $z\neq 0$, something that is not always explicit in the literature.}

We can also consider an ideal case, in which we could measure the flux of a source
in all wavelengths with a perfect detector ($S^X_\lambda(\lambda)=1,\;\forall\lambda$),
to define bolometric magnitudes,
\begin{equation}
m(t,z) := -2.5 \log \left (\frac{ \displaystyle\int^\infty_{\lambda=0}
f_\lambda(\lambda,t,z)
 d\lambda }{\displaystyle\int^\infty_{\lambda=0} g_\lambda(\lambda)
 d\lambda} \right ), 
\label{map}
\end{equation}
\begin{equation}
M(t) := -2.5 \log \left (\frac{ \displaystyle\int^\infty_{\lambda=0}
\frac{L_\lambda(\lambda,t)}{4 \pi
(10 pc)^2}
 d\lambda }{\displaystyle\int^\infty_{\lambda=0} g_\lambda(\lambda)
 d\lambda} \right ).
\label{Mab}
\end{equation}

{The distance to an astronomical object is directly related to its {bolometric 
magnitudes} through the quantity called \emph{distance modulus}:}
\begin{equation}
{\mu := m - M.}
\label{mbol}
\end{equation}

{As we will discuss in Section \ref{sec:dependence}, the observed spectrum of a source is modified with respect to 
its intrinsic one by the redshift, and therefore the radiation emitted in a given wavelength range in
the source's rest frame will be observed in a different range in the observer's frame. Also, since we simply cannot 
measure bolometric magnitudes, but only magnitudes in some {filters}, 
it is useful to express the distance modulus in terms of filter magnitudes, which requires the introduction of the 
so called $K$-correction $K_{XY}$ defined as}
\begin{equation}
{K_{XY} := m_Y - M_X - \mu .}
\label{keo}
\end{equation}

A full discussion of $K$-corrections and their applications for cosmology are left by the authors to another paper.

\section{Dependence of specific flux on redshift and distance} 
\label{sec:dependence}

It is important to note that even for a class of objects with the same intrinsic 
luminosity, which is approximately the case of SNe Ia (apart from the variations 
mentioned in section \ref{sec:introduction}), their observed fluxes (both specific and 
bolometric) will differ mainly due to the different redshifts and distances. 

{From (\ref{specific_flux}) we can see that, at a given time $t$ and at
a given wavelength $\lambda$, the specific flux can vary with distance $r$ to
the source, with redshift $z$, and with the functional form of the specific
luminosity $L_\lambda$. Considering SNe Ia as standard candles means that we
will assume all events to have the same specific luminosity. We know, however,
that there are variations in their luminosities that should be taken into
account and this will be considered in section \ref{sec:LC_standardization}. In the
present section we will study how an arbitrary observed spectrum differs from
the source's rest-frame spectrum, as we change, independently, the distance
$r$ and the redshift $z$. To that end, we advise the reader to refer now to
\ref{sec:basic_transformations}, where we graphically remind what happens to a function which is
subjected to certain simple transformations that will be relevant in the next
subsections.}

\subsection{Distance} 
\label{subsec:distance}

\begin{figure}[ht]
\center
\includegraphics[scale= 0.33]{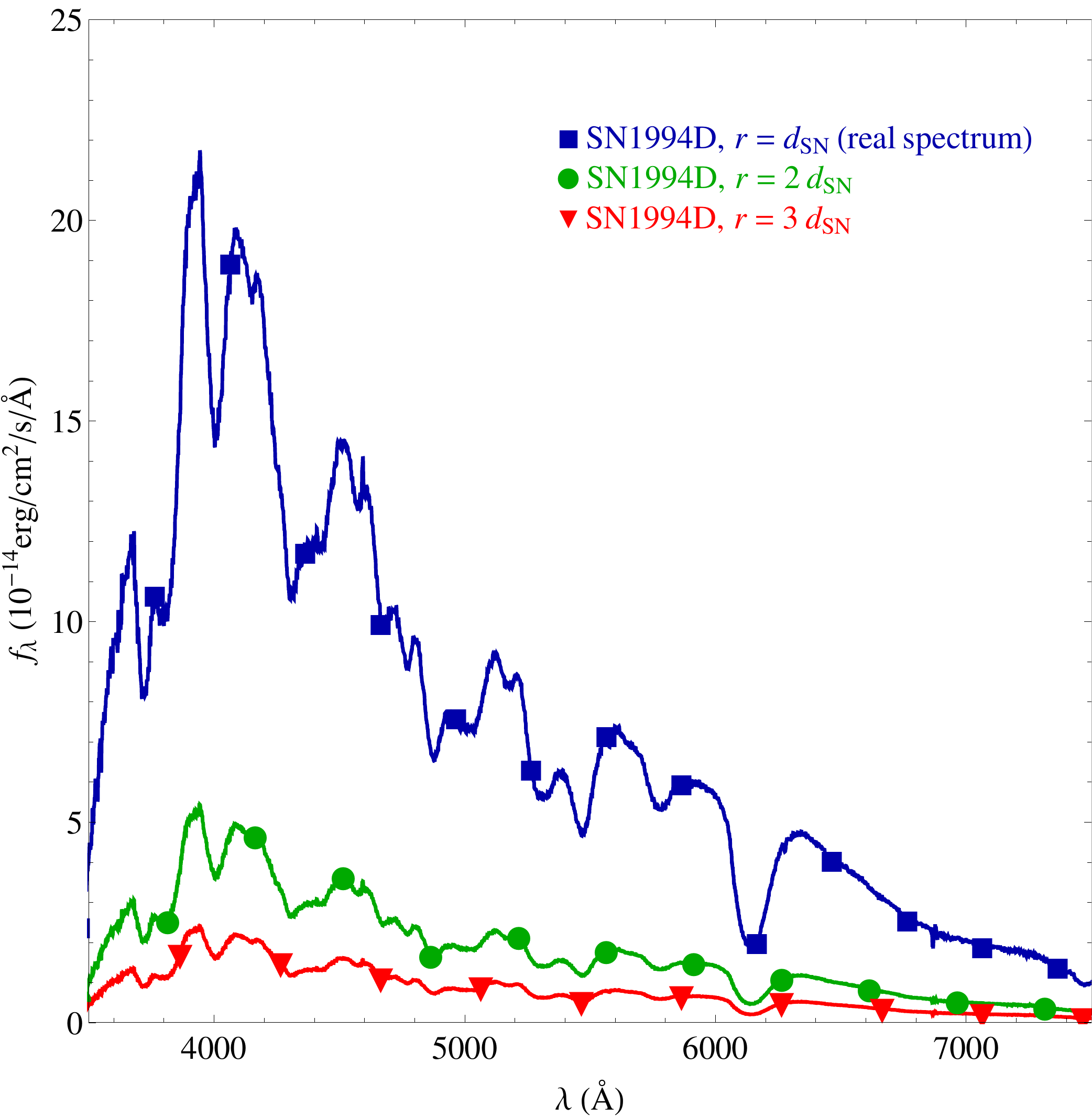}
\qquad\includegraphics[scale= 0.345]{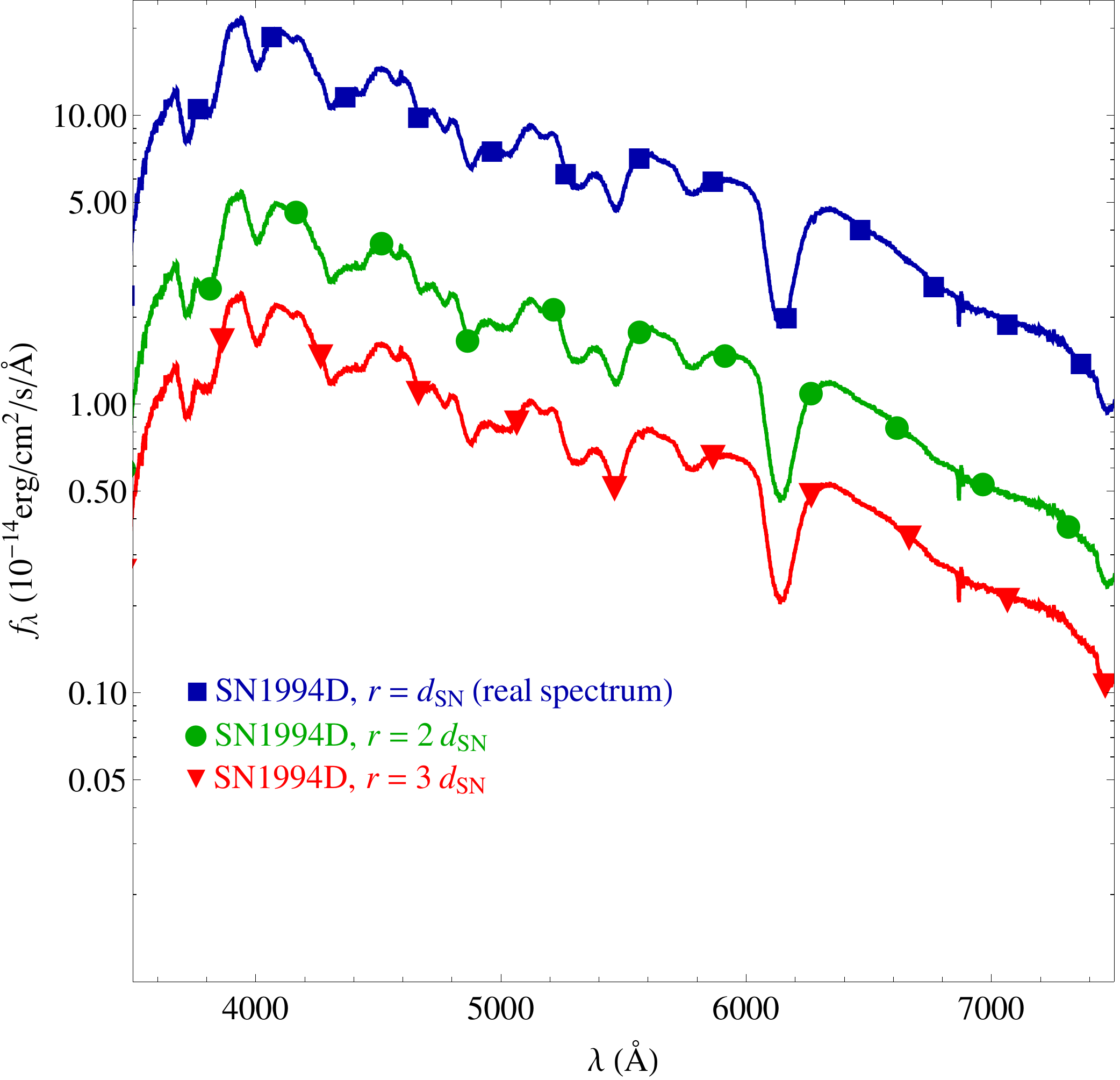}
\caption{Synthetic spectra simulating the effect of distance on the spectrum of 
SN Ia 1994D,
taken 2 days after maximum in the $B$ band, in linear (left panel) and 
logarithmic (right panel)
scales. The same spectrum was divided by different constants in order simulate 
different
distances (cf. (\ref{specific_flux})).}
\label{fig:distance}
\end{figure}

{Let us analyze first the simpler effect, the one arising from distance 
changes only. From (\ref{specific_flux}), we can see the dependence of the 
specific flux on the inverse square of the distance $r$. Thus, when 
\begin{equation}
 r\longmapsto r'=cr\,\quad (c=\mbox{const.})\,,
\end{equation}
all other independent variables held constant, we have that
\begin{equation}
 f_\lambda(\lambda,t,r,z,L_\lambda)\longmapsto {f'_\lambda(\lambda, t, r', z, 
L_\lambda)} = c^{-2}\,f_\lambda(\lambda, t, r, z,
L_\lambda)\,.
\end{equation}
Therefore, in a graph of the spectrum, as shown in figure \ref{fig:distance} for 
SN1994D, we employ, in a linear scale (left panel), the vertical distortion of 
(\ref{vert_dist}) and, in a logarithmic scale (right panel), the vertical 
translation of (\ref{vert_trans}). The effect on the spectrum of a pure 
change only in distance is manifest in the logarithmic scale, where the rigid 
vertical translation is obvious.} 

\subsection{Redshift}
\label{subsec:redshift}

\begin{figure}[ht]
\center
\includegraphics[width= 0.45\textwidth]{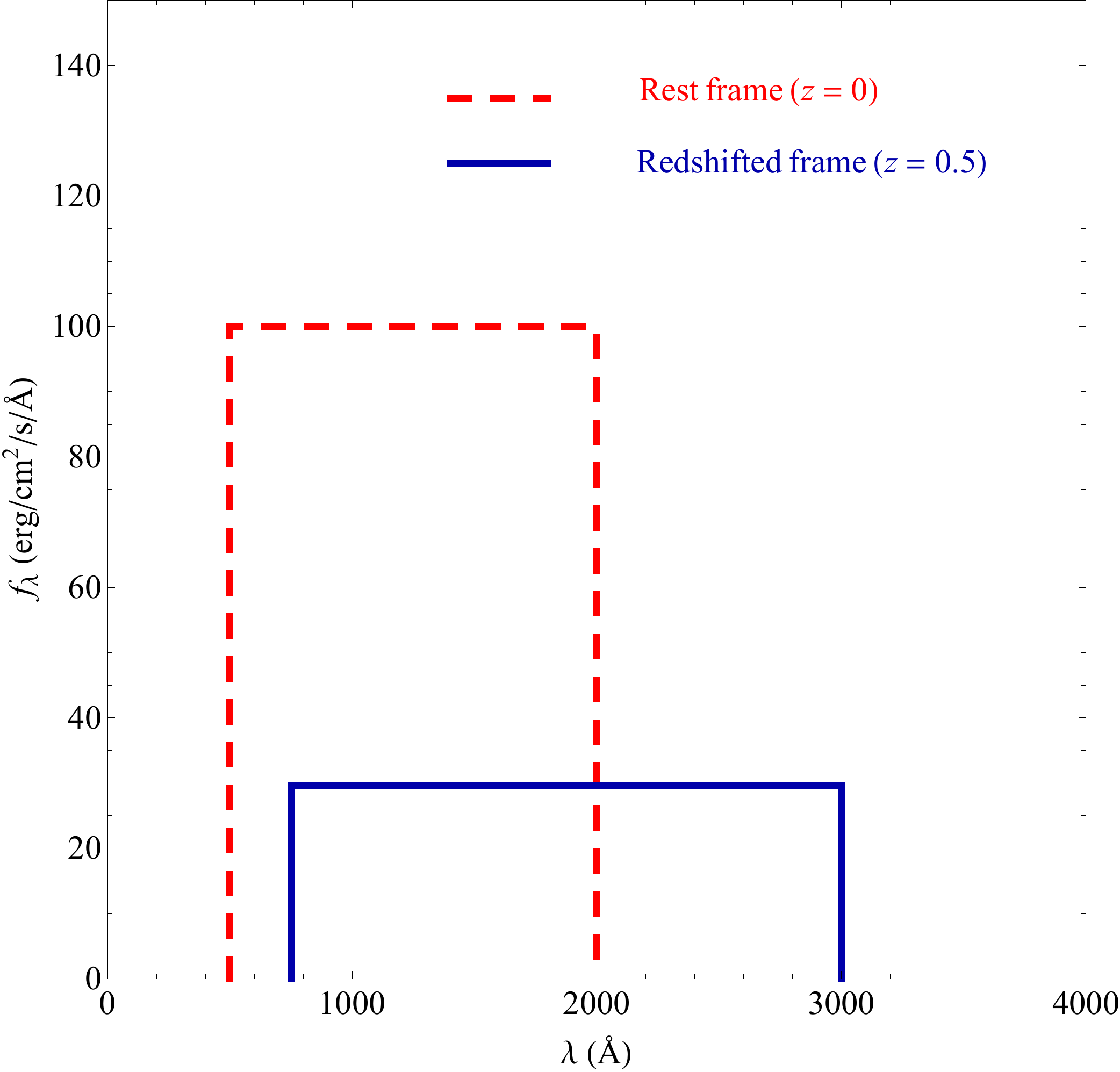}
\qquad\includegraphics[width = 0.45\textwidth]{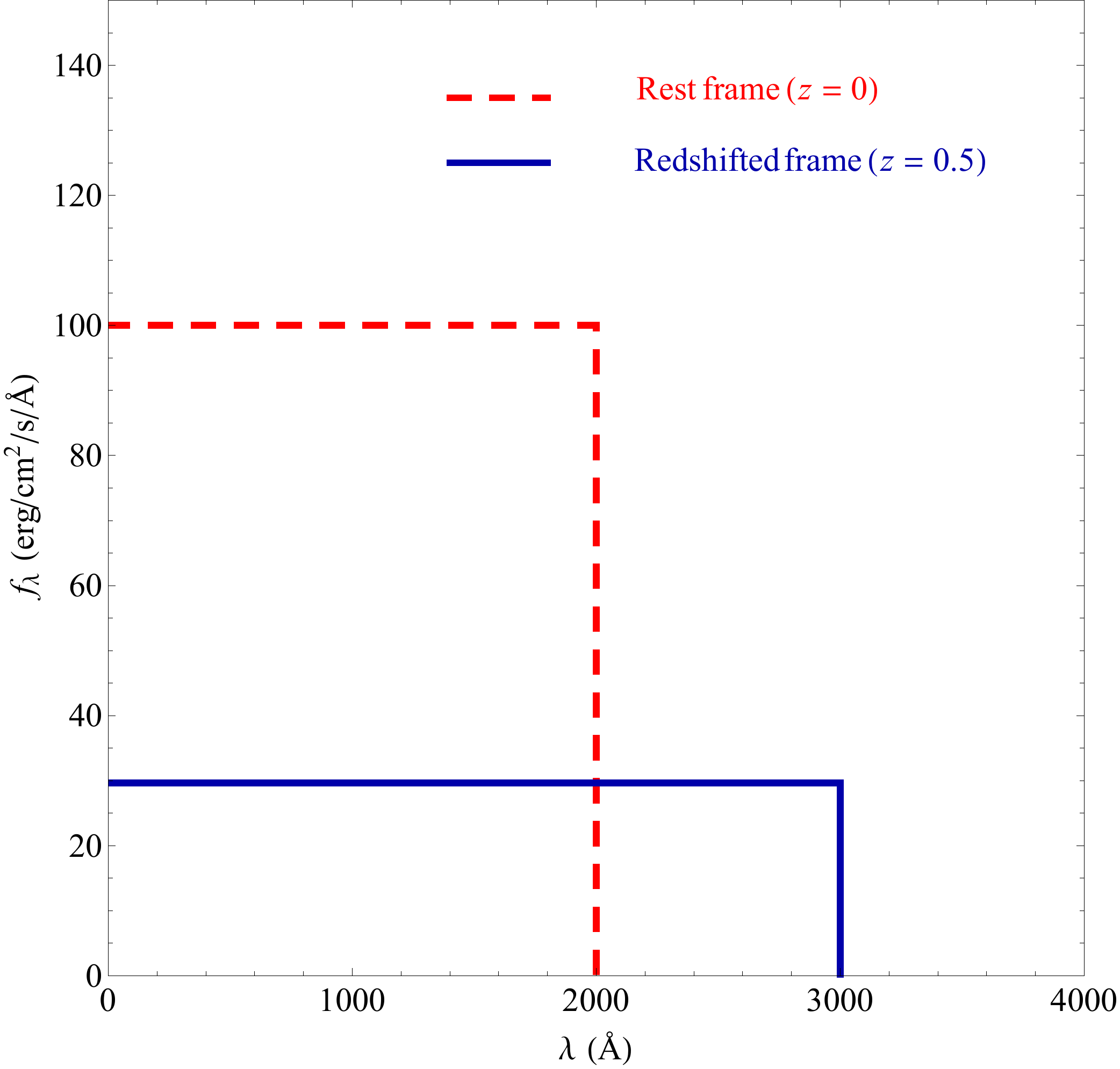}
\caption{Effect of a variation in redshift on two top-hat spectra. The blue
curve is the rest-frame spectrum ($z=0$) and the red one is same spectrum
at a redshift $z=0.5$, in the observer's frame.}
\label{fig:redshift}
\end{figure}

Let us analyze now the effect of the redshift, related to the relative motion
between source and observer. Again, from (\ref{specific_flux}), we can see the dependence of
the specific flux on the inverse cube of $(1+z)$ and also modifying
explicitly the independent variables $\lambda$, and $t$ by factors of
$1/(1+z)$. Thus, when
\begin{equation}
 1+z\longmapsto 1+z'=c(1+z)\,\quad (c=\mbox{const.})\,,
\end{equation}
all other independent variables held constant, we have that
\begin{equation}
 f_\lambda(\lambda,t,r,z,L_\lambda) \longmapsto
{f'_\lambda(\lambda,t,r,z',L_\lambda)} = c^{-3}\,f_\lambda\!\left( \frac{\lambda}{c},
\frac{t}{c},r,z,L_\lambda \right) \,.
\label{redshift_transformation}
\end{equation}
Of course, referring to the Appendix, we see that this transformation of
the specific flux involves the composition of a vertical distortion,
(\ref{vert_dist}), and a horizontal distortion, (\ref{hor_dist}).
To get a handle on it more intuitively, let us choose $z=0$ so that the
former equation will provide the redshifted spectrum from the rest-frame one:
\begin{equation}
 {f'_\lambda(\lambda,t,r,z',L_\lambda)} =
\frac{1}{(1+z')^3}\,f_\lambda\!\left(\,
\frac{1}{1+z'}\lambda, \frac{1}{1+z'}t, r, z=0, L_\lambda \,\right)\,,
\label{rest_frame2redshifted_spectrum}
\end{equation}
or vice versa, the rest-frame spectrum from the redshifted one:
\begin{equation}
 f_\lambda(\lambda,t,r,z=0,L_\lambda) =
(1+z')^3\,{
f'_\lambda \! \left( \,(1+z')\lambda , (1+z')t , r , z' ,
L_\lambda\, \right)}\,.
\end{equation}
{Now, to illustrate this redshifting effect in a most pristine situation},
we apply (\ref{rest_frame2redshifted_spectrum}) to a top-hat function. The
result is shown in figure~\ref{fig:redshift}.
In the left panel, we show that the total qualitative effect of {the
redshift} is: {(i) a vertical {squeezing}, due to the {$1/(1+z')^3$} 
pre-factor, and (ii) a horizontal stretch
caused by the rescaling {$\lambda \longmapsto \lambda/(1+z')$} in the first 
argument of $f_{\lambda}$. From
this panel, the reader could naively be induced to regard the displacement 
towards greater wavelengths  as a third, independent, effect; however, as can 
be seen from the right panel of figure~\ref{fig:redshift}, such a displacement is 
in fact also due to the horizontal stretch, which leaves the vertical $y$-axis 
($\lambda=0$) fixed (cf. (\ref{hor_dist}) and right lower panel of
figure~\ref{fig:basic_function_transformations}).}

{In figure~\ref{fig:d}, left panel, we showed observed spectra of three SNe
Ia. In its right panel, we now show the corresponding rest-frame ($z=0$)
spectra. We can notice the small horizontal displacement of the spectral lines
(blueshifted, towards the left) but it is not possible to visualize the vertical
displacement (upwards) due to the low values of the redshift involved.} We can
also see that, even after the redshift correction, the spectra do not coincide
and this is because each SN is at a different distance from us.

\begin{figure}[ht]
\center
\includegraphics[scale= 0.34]{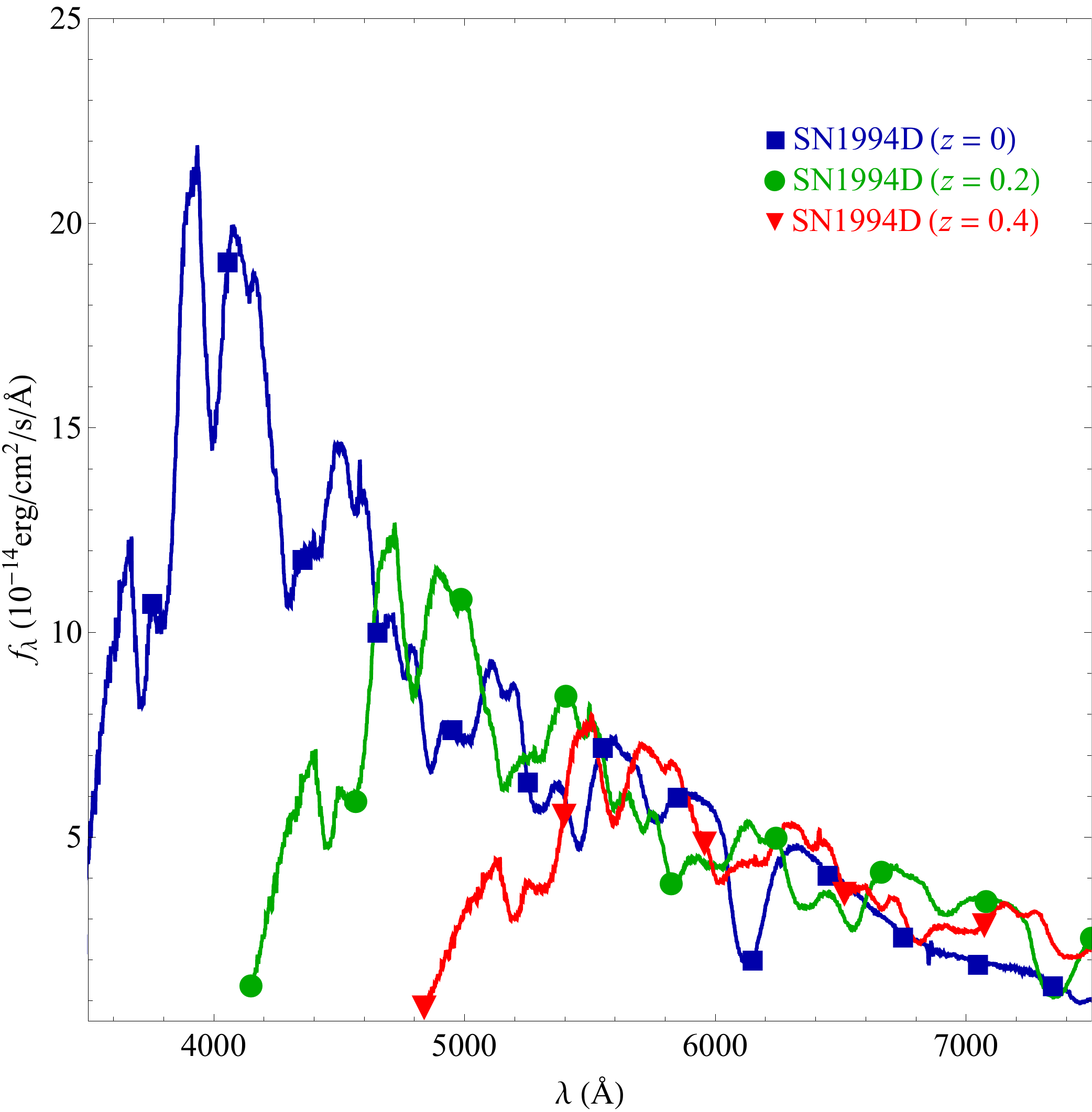}
\qquad\includegraphics[scale= 0.35]{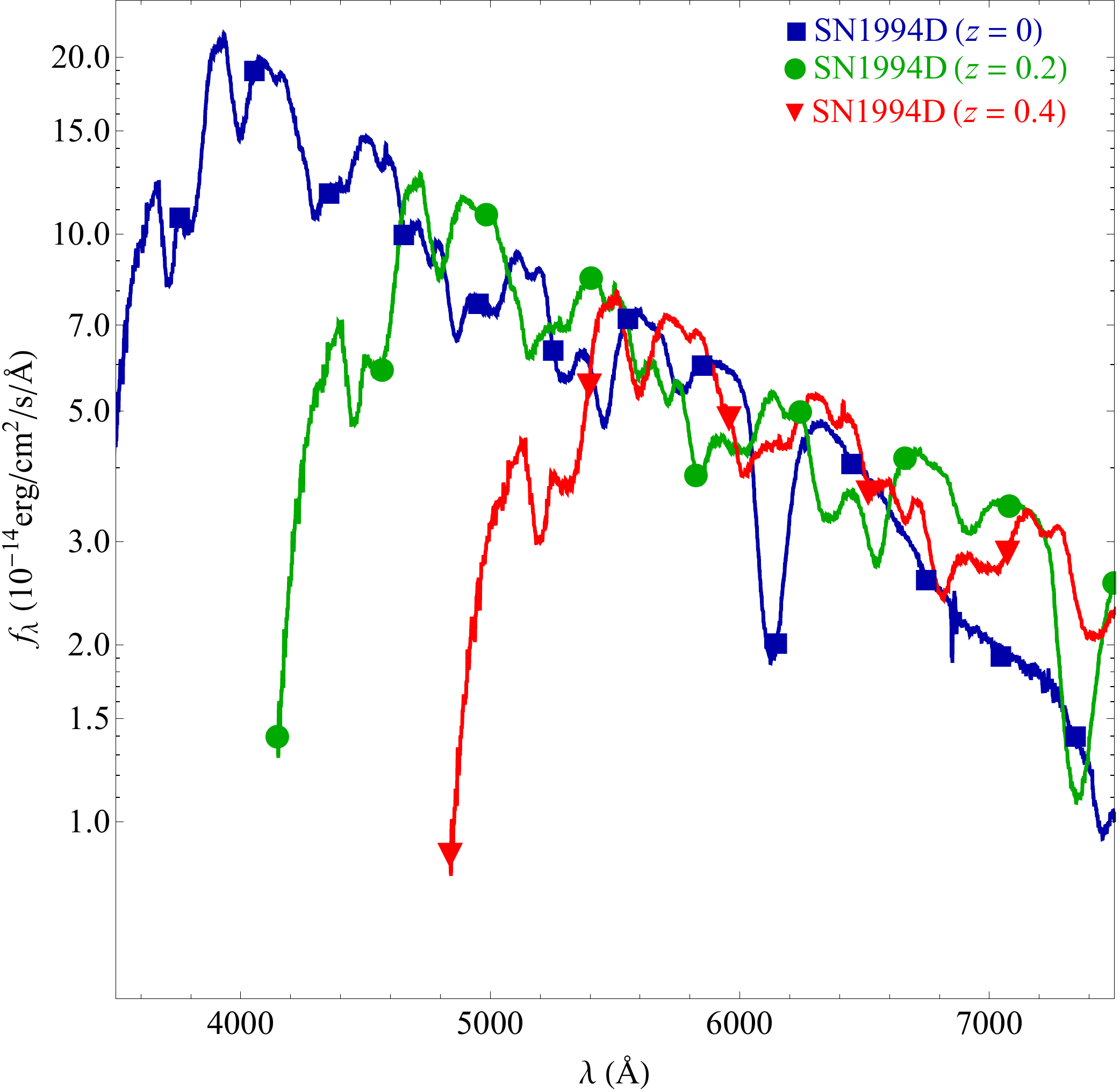}
\caption{Synthetic spectra simulating the effect of redshift on the spectrum of
SN Ia 1994D, taken 2 days after maximum in $B$ band, for different values of
redshift. We use a linear scale in the left panel and a logarithmic one in the
right panel.}
\label{fig:z}
\end{figure}

{To explicitly reveal the redshifting effect on the spectrum of a concrete 
SN Ia, we show, in figure~\ref{fig:z}, three spectra of SN 1994D, the rest-frame 
one and two other (artificial) high redshift ones (left panel). In particular, the 
effect of the
pre-factor $1/(1+z')^3$ in (\ref{rest_frame2redshifted_spectrum}) can be 
best viewed using a logarithmic scale (right panel), in which it becomes a 
simple vertical translation (cf. (\ref{vert_trans}) and left upper panel of
figure~\ref{fig:basic_function_transformations}).}

\section{Light curve standardization} 
\label{sec:LC_standardization}

Although source-frame SNe Ia light curves are very similar, they are not 
identical. In this section we will show that it is possible to make them even
more similar by applying some simple operations, which are dubbed
standardization, {and we will apply this procedure 
to a sample of real type Ia SNe. The process of standardization became possible
after the discovery that intrinsically brighter SNe (at $B$ band maximum) were also the ones with wider
light curves.\cite{Phillips93,Hamuy96} Such a correlation rendered it possible to determine 
if a given SN was brighter (fainter) than another one either because it was closer (further) 
or because it was intrinsically brighter (dimmer), just by looking at their light curves.}

{The data used in this work are publically
available,\cite{CSPWebsite} and constitute the sample of 85
low redshift SNe Ia observed by the Carnegie Supernova Project (CSP).\cite{Contreras10,Stritzinger11} Motivated by the higher uniformity of SNe Ia
in the infra-red band, one of the main goals of that project was
to obtain particularly well sampled and well characterized light curves both in
optical and near-infrared bands, which should improve the efficiency
of the standardization process. We restricted ourselves to the subsample
of only Branch-normal SNe Ia, which reduced the number of events to 71.} 
The corrections that we will present here were originally done simultaneously 
{through a single fit that yields all the correction factors for each SN (cf. Goldhaber \emph{et al.} \cite{Goldhaber01})}; however
we chose to implement them step by step in order to make clear the role of each
one in the final result. 

\begin{figure}[ht]
\center
\includegraphics[scale= 0.7]{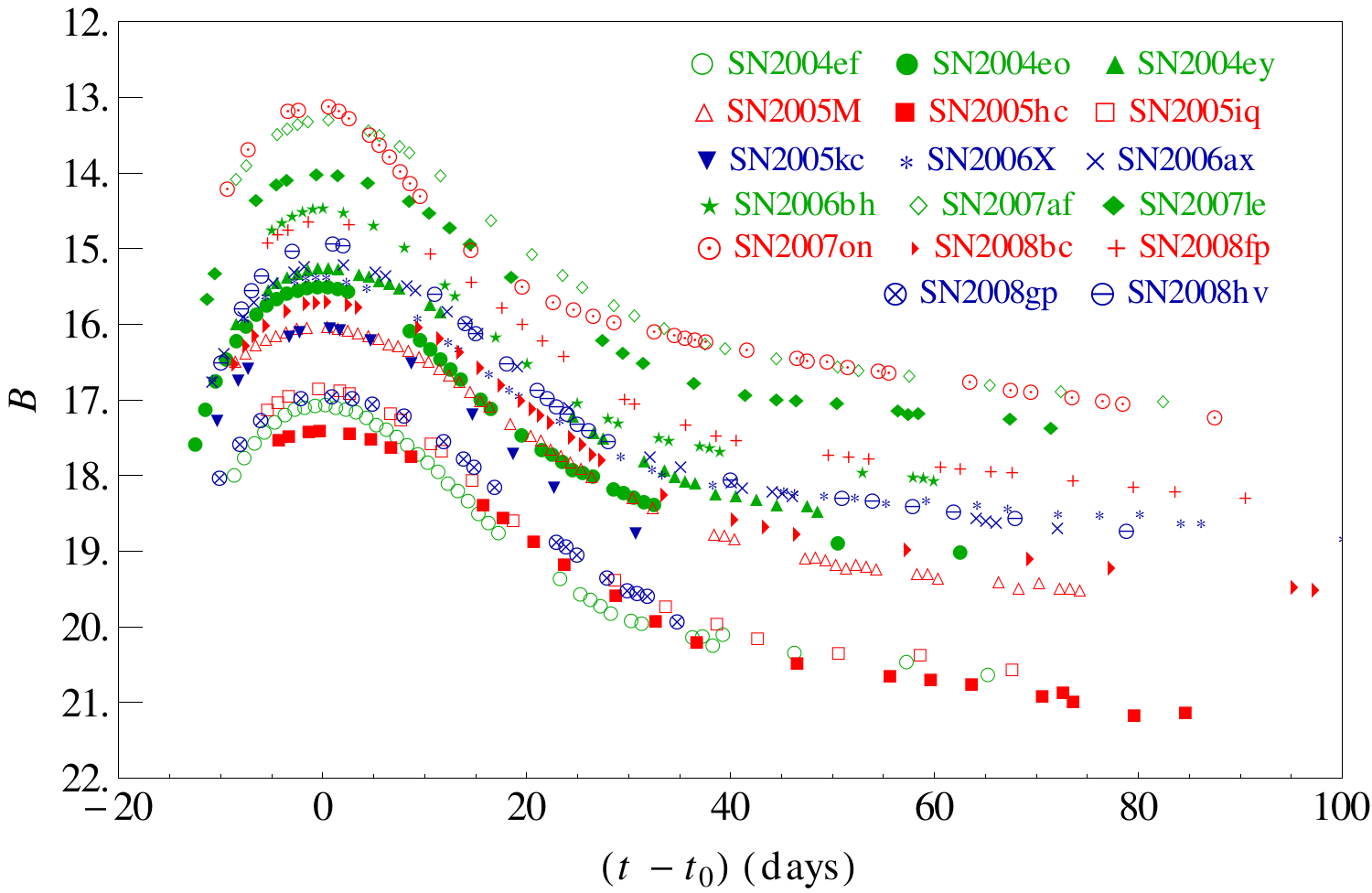}
\caption{{Apparent magnitude \emph{B} band light curves of the 17 SNe Ia in our subsample after the time axis offset correction (cf. subsection \ref{subsec:TimeAxisCorr}).}}
\label{min}
\end{figure}

\subsection{{Time axis offset correction}}
\label{subsec:TimeAxisCorr}
\setcounter{footnote}{0}
{In the CSP light curve data, the epoch is expressed in Modified Julian Date
(MJD). In order to compare them in a single plot, we need to define a common
time scale $t-t_0$\footnote{The time scale $t-t_0$ is commonly called \textit{phase}.}, where $t_0$ is the epoch of maximum flux,
traditionally considered in $B$ band. We wrote a simple code to obtain $t_0$ for
each supernova in our subsample. Unfortunately, some of them were observed only
after $B$ band maximum and were thus excluded from our subsample, which reduced
considerably the number of SNe in the final subsample.} {In fact, we
required our code to keep only the SNe that presented at least 3 observations
taken before maximum flux and at least one observation taken after 30 days from
maximum flux (the reason for this restriction will become clear in section
\ref{subsec:StretchCorr}). This left us with a 
subsample of 17 SNe, whose names and redshifts are listed in table 
\ref{tab:SNlist},} {and whose time-offset-corrected light curves can be
seen in figure \ref{min}.} 

\begin{table}
\centering
\caption{Names, CMB-centric redshifts and stretch factors (see section
\ref{subsec:StretchCorr}) for all 17 SNe Ia in the final subsample used to
generate our simple light curve
template.}
\footnotesize
 \begin{tabular*}{\textwidth}{@{}l*{15}{@{\extracolsep{0pt plus12pt}}l}}
 \br
 & SN & $z_{\rm CMB}$ & $s$ & $s_{_G}$\\
 \mr
 1 & 2004ef & 0.0298 & 0.89 & 0.81\\
 2 & 2004eo & 0.0147 & 0.87 & 0.88\\
 3 & 2004ey & 0.0146 & 1.14 & 1.00\\
 4 & 2005M & 0.0230 & 1.15 & 1.11\\
 5 & 2005hc & 0.0450 & 1.14 & 1.10\\
 6 & 2005iq & 0.0329 & 0.93 & 0.89\\
 7 & 2005kc & 0.0139 & 0.98 & 0.92\\
 8 & 2006X & 0.0063 & 1.01 & 0.93\\
 9 & 2006ax & 0.0179 & 1.12 & 0.98\\
 10 & 2006bh & 0.0105 & 0.86 & 0.82\\
 11 & 2007af & 0.0063 & 1.01 & 0.94\\
 12 & 2007le & 0.0055 & 1.12 & 0.97\\
 13 & 2007on & 0.0062 & 0.62 & 0.70\\
 14 & 2008bc & 0.0157 & 1.20 & 1.03\\
 15 & 2008fp & 0.0063 & 1.18 & 1.06\\
 16 & 2008gp & 0.0328 & 1.07 & 0.98\\
 17 & 2008hv & 0.0136 & 0.97 & 0.88\\
 \br
 \end{tabular*}\\
 \label{tab:SNlist}
\end{table}
\normalsize

\subsection{{Distance and redshift corrections}}
\label{subsec:distance_redshift_corrections}

{In order to properly standardize the light curves, we need to correct them 
for extrinsic effects. As we have seen in section~\ref{sec:dependence}, two of 
them can be easily taken account of: distance and redshift. The latter entails 
a change of the time scale and an offset to the magnitude (or change of the 
flux normalization) whereas the former implies a simple offset to the 
magnitude. Thus the correction for both effects amounts to:}
\begin{enumerate}
 \item {a (horizontal) 
dilation, cf. (\ref{hor_dist}), of the time axis such that}
\begin{equation}
\frac{\Delta t_o}{\Delta t_e} = 1+z,
\end{equation}
where $\Delta t_o$ is a time interval in the observer's frame and $\Delta t_e$ 
is the
corresponding interval in the source's rest frame.
 \item {a (vertical) rigid translation, cf. (\ref{vert_trans}), of the magnitude axis.} 
\end{enumerate}

{Notice that after the rigid vertical translations to correct for the redshift and distance, 
it is possible that the {peaks of the} light curves still do not coincide, since there can be 
 absolute magnitude differences {among}  them. So, in order to make 
{the peaks}  coincide, 
 a third vertical rigid translation is still needed. In our case, we {do}  not know the distances
to the SNe in our sample, so what we actually did was to evaluate the {peak} magnitude's mean, and 
displace the light curves in order to make their magnitudes {match}  this mean. This operation 
accounts for the {rigid} vertical translations due to both the redshift and the distance corrections, and also to 
{a}  third rigid translation to correct for {other}  differences in absolute magnitude. 
 }

\begin{figure}[ht]
\center
\includegraphics[scale= 0.7]{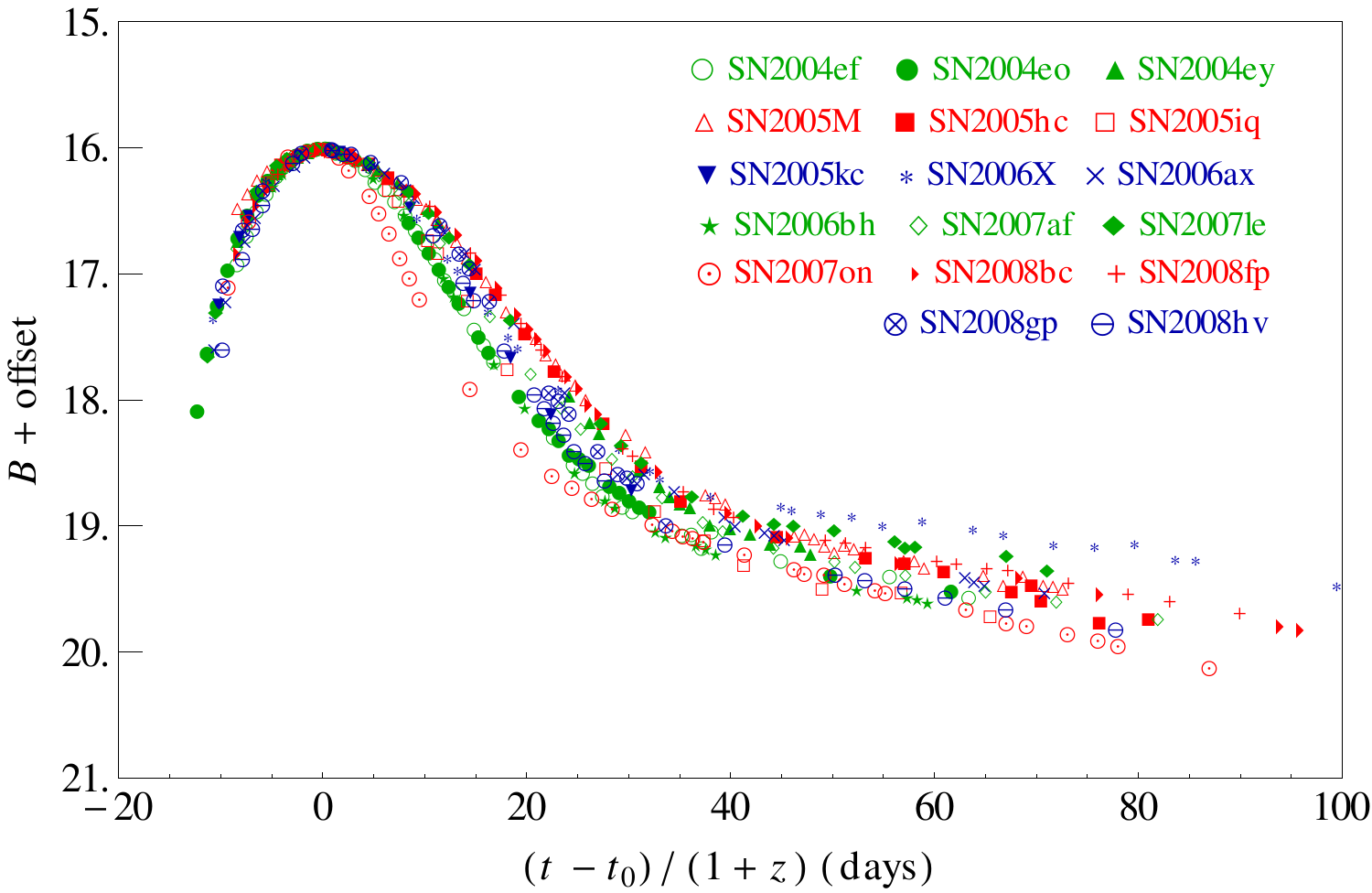}
\caption{{Apparent magnitude \emph{B} band light curves $+$ offset for the 17 SNe in our subsample after the time axis offset (cf. subsection~\ref{subsec:TimeAxisCorr}) and distance and redshift corrections (cf. sunsection \ref{subsec:distance_redshift_corrections}).}}
\label{y}
\end{figure}

{The resulting distance- and redshift-corrected light curves of our 
subsample are shown in figure~\ref{y}}. In order to display all 
SNe in their rest frame time, notice that we have chosen to change 
the $x$-axis from $t-t_0$ to $(t-t_0)/(1+z)$. Because of this, a little 
bit of care must be taken when comparing figure~\ref{y} and the following 
figures in this section to the results presented in section~\ref{sec:dependence} 
and \ref{sec:basic_transformations}, where we are keeping the $x$-axis 
unchanged before and after a given transformation.
 
\subsection{The stretch correction}
\label{subsec:StretchCorr}

\begin{figure}[ht]
\center
\includegraphics[scale= 0.7]{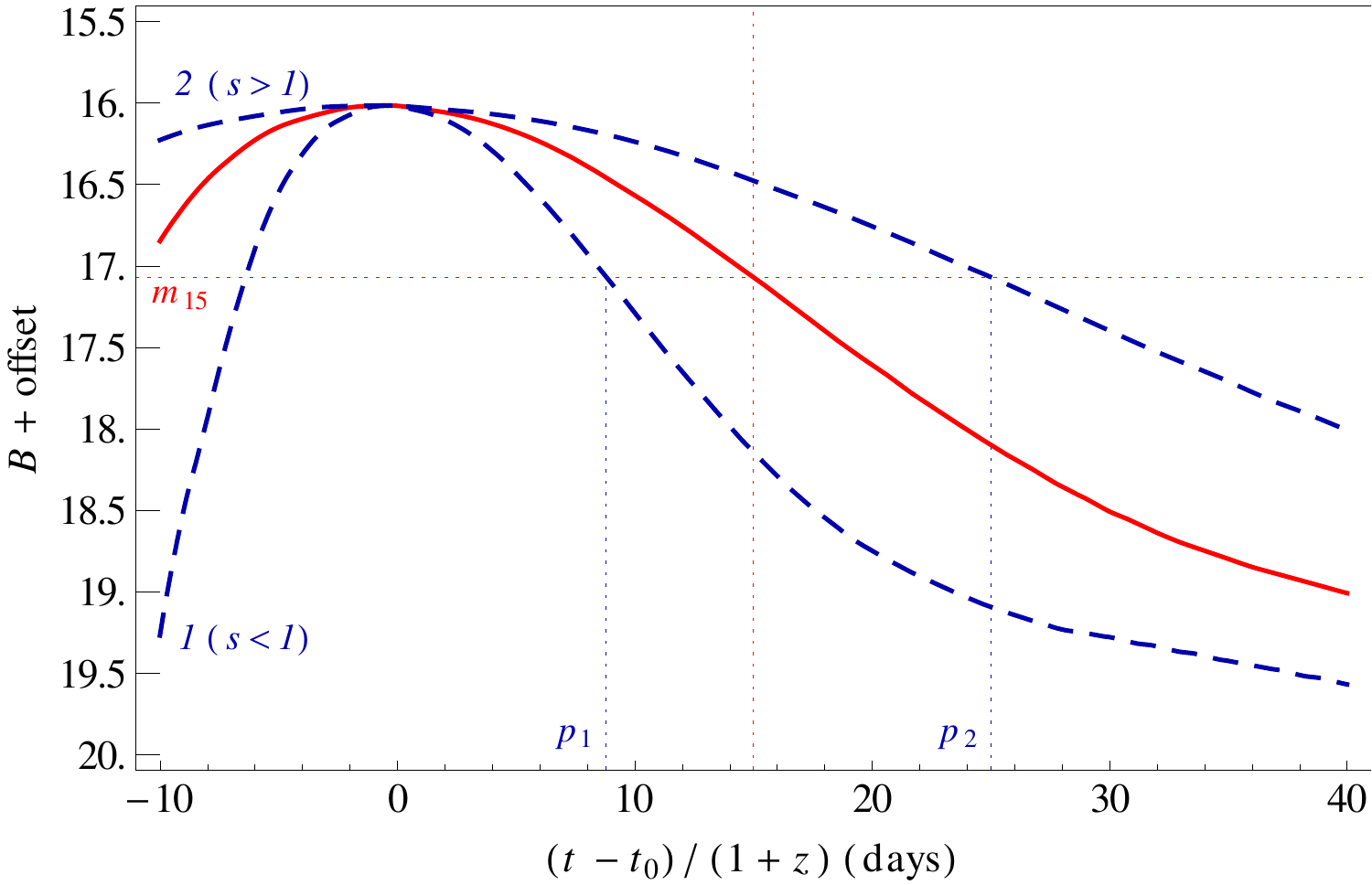}
\caption{Schematics for stretch calculation. The red solid curve is the fiducial light curve, for which { $s=1$, by definition}. The red dotted horizontal line indicates the position of $m_{15}$ in the $y$-axis, which corresponds to the point in the fiducial curve where $(t-t_0)/(1+z)=15$ days, marked with the {red dotted vertical line}. The {blue dashed curves} are two other fictitious light curves, upon which we want to apply the stretch correction. The two {blue dotted lines} show the values of $p$ for each curve, and the resulting range of the stretch factor is indicated near each curve.}
\label{cstretch}
\end{figure}

The stretch parameter $s$ is related to the width of the light curve, i.e. it 
measures how fast the supernova's flux decreases {(or its magnitude 
increases).\cite{Perlmutter99,Goldhaber01}} 
In order to calculate the stretch, we need to adopt a fiducial curve which, in 
our case, 
was chosen to be simply the mean of all curves in the sample, and assign the 
value
$s=1$ to it. A curve that declines slower (faster) than the fiducial one will
have 
$s>1$ ($s<1$). After the corrections described in subsections \ref{subsec:TimeAxisCorr} and
\ref{subsec:distance_redshift_corrections} all curves coincide at the $B$ band maximum but not necessarily at any other point.
The stretch correction is designed so that the curves also coincide at 15 source frame days after $B$ band maximum.
 {We show a sketch of this procedure in 
figure \ref{cstretch}, in which we use a fiducial (red curve) and
two ficticious light curves, $1$ and $2$, in blue.} 

To obtain the stretch we need to
solve {for $p_i= (t-t_{0,i})/(1+z_i)$ from the following equation}
\begin{equation}
f_i(p_i)=\overline{m}_{15},
\end{equation} 
where $f_i$ is an interpolating function (in our case a spline) for the 
{$i$-th SN Ia $B$ band light curve}, and $\overline{m}_{15}$ is the value 
of 
$B$ (+ offset) of the mean light curve at {$p_i=15$ days}. 
Thus, 
the stretch can be written as
\begin{equation}
s_i=\frac{p_i}{15{\mbox{ days}}}.
\end{equation}

\noindent We can then divide all phases of a supernova by the obtained stretch 
so the 
curves coincide at
phase 15 days.

As mentioned earlier, the width difference in the light curves is associated to their intrinsic 
brightness (broader $\leftrightarrow$ brighter). 
When we correct 
for the stretch
we are compensating the differences in intrinsic brightness between the 
supernovae. 

\begin{figure}[ht]
\center
\includegraphics[scale= 0.7]{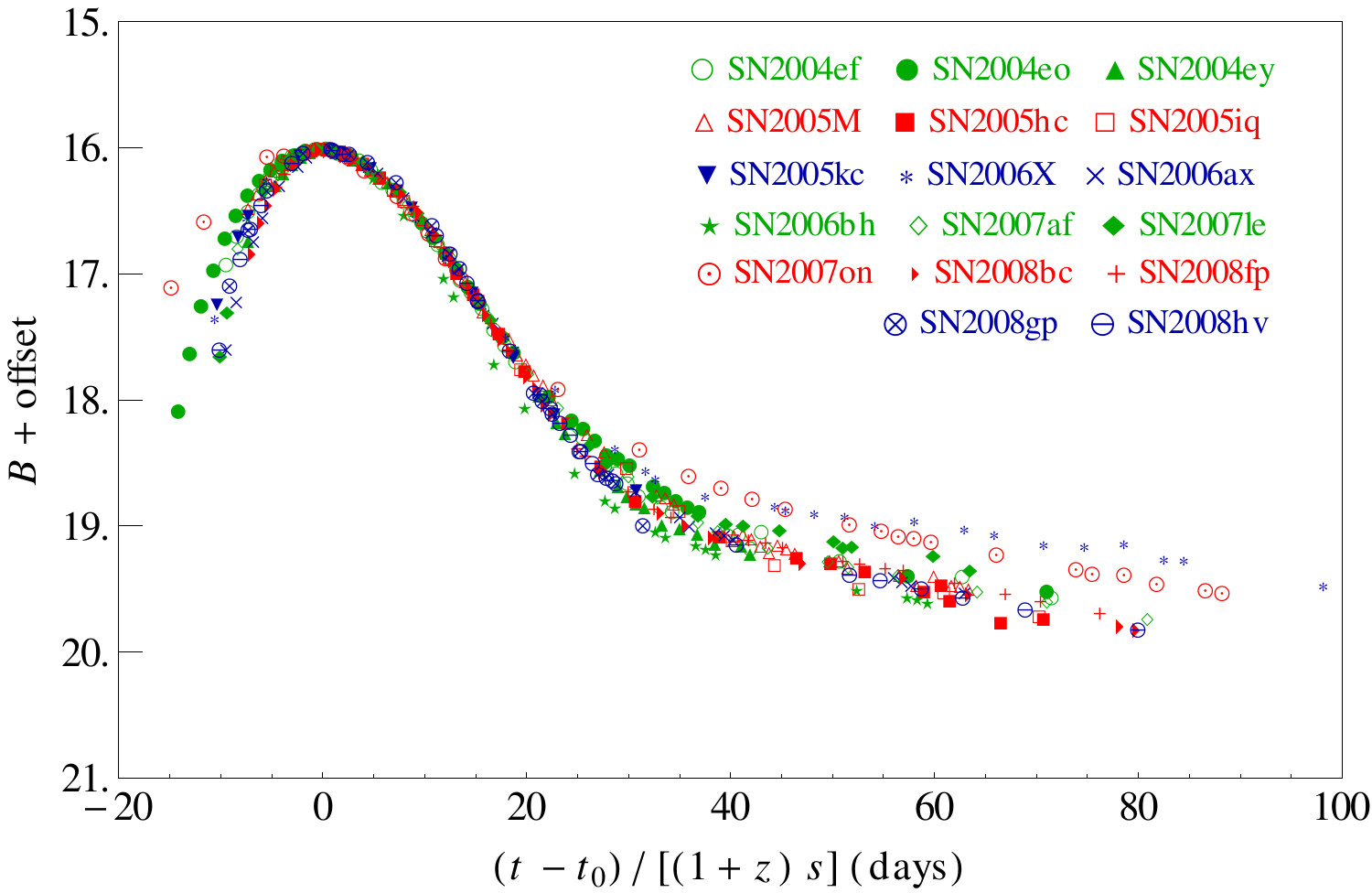}
\caption{{Apparent magnitude $B$ band light curves $+$ offset of the 17 SNe Ia in our subsample after the time axis offset (cf. subsection~\ref{subsec:TimeAxisCorr}), distance and redshift (cf. subsection \ref{subsec:distance_redshift_corrections}) and stretch corrections (cf. subsection \ref{subsec:StretchCorr}).} Note that the curves
still present {some} dispersion, since the recipe imposes coincidence 
{only} at two points: {$(t-t_0)/[(1+z)s]=0,\;15$ days}.}
\label{s}
\end{figure}

The result of the application of the stretch procedure to our sample can 
be seen in figure \ref{s}. {Table \ref{tab:SNlist} shows the values of the stretch $s$ 
found for each SN, according to our procedure. For comparison, we also show the values of the same parameter, now dubbed $s_{_G}$, found using 
the method described in Goldhaber \textit{et al.},\cite{Goldhaber01} where the corrections are all done simultaneously
and the fiducial curve is different from ours. We can see that the results obtained with our 
step-by-step method agree quite well with the ones obtained with this more sophisticated fitting 
recipe.}

\subsection{{The resulting template}}

\begin{figure}[ht]
\center
\includegraphics[scale= 0.45]{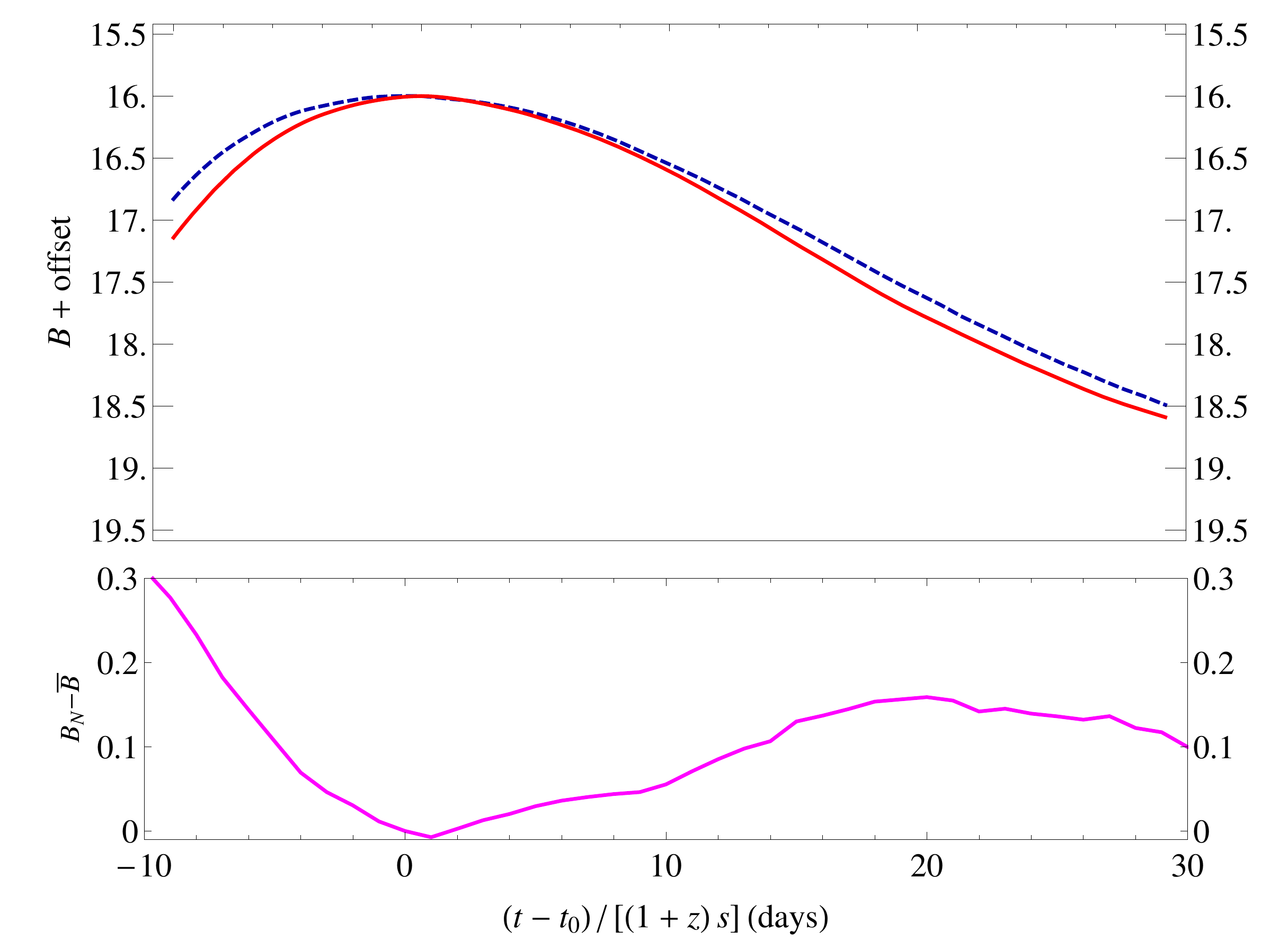}
\caption{{Comparison between our $B$ band light curve template and Nugent's 
one.\cite{Nugent_templates} Upper panel: $B$ band rest-frame magnitude 
(arbitrarily normalized) versus rest-frame stretched phase for our 
template (red, solid curve) and for Nugent's one (blue, dashed curve). Lower 
panel: {discrepancy between our template and Nugent's one}.}}
\label{template}
\end{figure}

After we apply all the corrections discussed above we can construct what we can 
call a
``rudimentary" template, a simple mean of all corrected curves, that can be 
compared to {the Nugent template,}
one of the most used in the literature.\cite{Nugent_templates} We show
the
comparison
in figure \ref{template}. {In this comparison we can not use the relative discrepancy
between these curves (\,$(B_N-\bar{B})/B_N$ or $(B_N-\bar{B})/\bar{B}$\,) because the 
normalizations of both are \emph{arbitrary}. We can, on the other hand, compare
the absolute difference shown in figure \ref{template} (lower panel) to the range of the
Nugent template in the depicted interval ($[-10,30]$), for instance, which is $2.5$, and the discrepancy thus calculated is always less than $12\%$.} We can see that, despite the simplified analysis 
performed here,
our curve looks quite similar to the template, which {shows the consistency 
between our template and the Nugent one}. 

\begin{figure}[ht]
\center
\includegraphics[scale= 0.40]{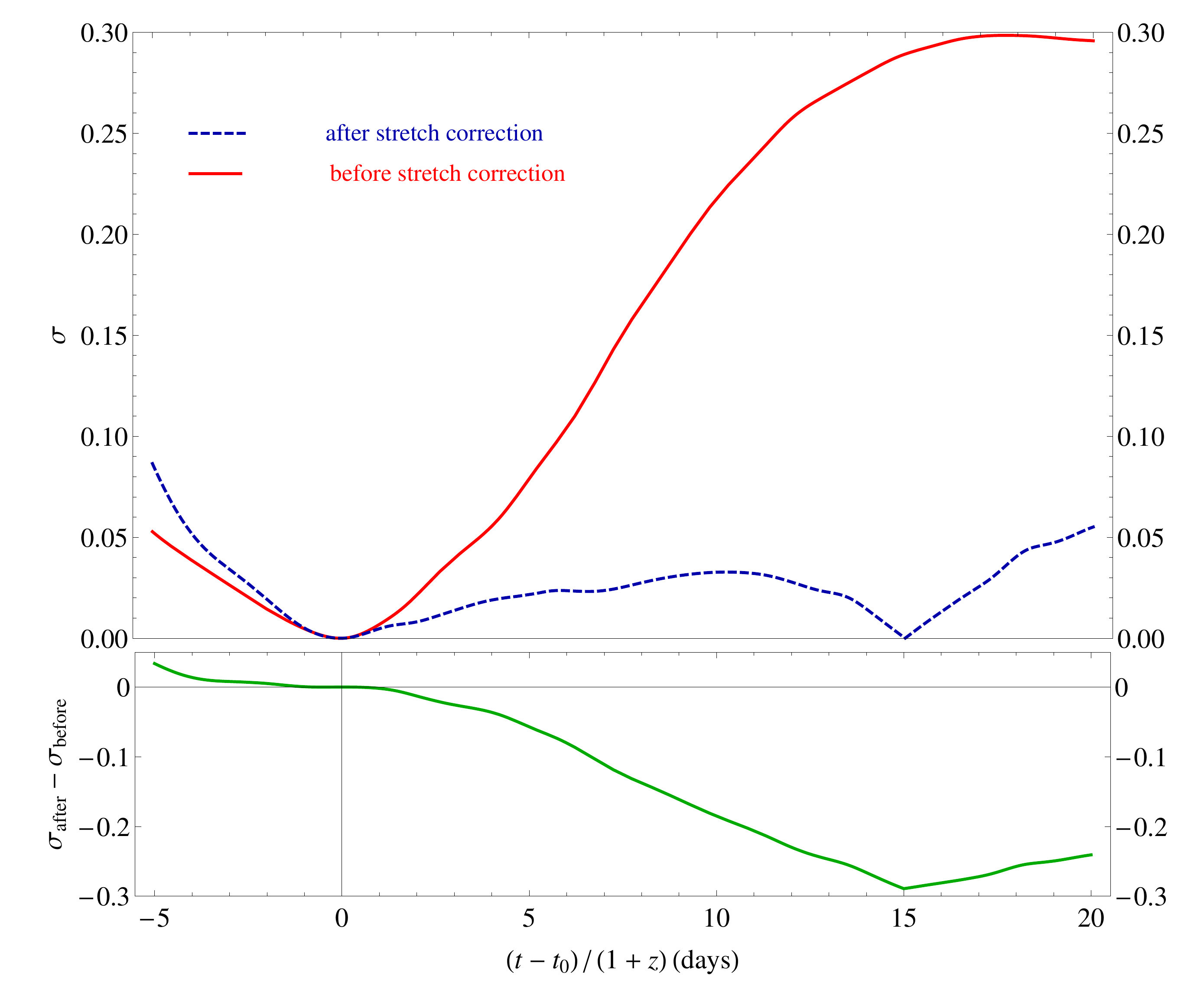}
\caption{{The role of the stretch correction in diminishing the dispersion. 
Upper panel: $B$ band light curve standard deviation for the 17 SNe Ia 
after all corrections (blue, dashed curve), and with all but the stretch 
correction (red, solid curve). Lower panel: discrepancy between the 
standard deviations in magnitude of our subsample after all corrections 
(including the stretch one) and before the (last) stretch correction.}}
\label{sigma}
\end{figure}

We show in figure \ref{sigma} the 
standard deviation of our sample before and after
the stretch correction. {Again, we can not use relative discrepancies in the whole time interval to compare these curves
because both the standard deviations are zero at $t-t_0=0$ by construction (see subsection \ref{subsec:distance_redshift_corrections}) and the stretch corrected one is also null at $(t-t_0)/[s(1+z)]=15$. Nevertheless, the overall decreasing in the
dispersion after the maximum is clear from figure \ref{sigma} . Since such gain is obtained through a simple linear transformation, with only one parameter, we can argue that it} {reflects the homogeneity of the light curves 
in our sample}

{The reader might note that the discrepancy between the standard deviation 
after the stretch correction (in figure \ref{sigma}) 
is greater before maximum. This feature reflects the fact that 
the dispersion of the SNe Ia is smaller before the maximum (see Hayden 
\textit{et al.}\cite{Hayden10}). The stretch is defined to
decrease the dispersion after maximum but it is applied to the whole light 
curve through (\ref{hor_dist}) therefore, since the curves are more uniform before 
the maximum, when we multiply their arguments by different numbers the net result is 
an increasing of the dispersion in this interval. }

\section{Conclusion} 
\label{sec:conclusion}

This article had two main aims: (i) presenting and clarifying some 
fundamental concepts and results related to the cosmological use of SNe Ia, and 
(ii) building a simple SN Ia light curve template. 

The first aim led us to introduce, in section~\ref{sec:fundamental_quantities}, the 
specific flux or spectral energy distribution as the principal quantity characterizing 
the class of transient SNe Ia, and the corresponding projections (spectra and specific 
light curves). In section~\ref{sec:dependence}, we studied in particular its 
dependence on distance and redshift and the consequent impact on the observed 
fluxes or magnitudes. 

To comply with the second aim cited above, in 
section~\ref{sec:LC_standardization}, we built our naive light curve template, 
for didactic purposes, through a simplified version of the original stretch 
procedure: we performed the determination of the three parameters of the method (the overall 
normalization of the light curve, the epoch of maximum flux in $B$ band and the
stretch itself) 
{separately}, instead of the simultaneous fit described
in Goldhaber \textit{et al.}.\cite{Goldhaber01} We finally constructed a mean light curve after the
application of the method and compared it to a light curve template much used in the 
literature,\cite{Nugent_templates} {showing that our simplified method is able to produce 
a template very similar to it. In fact, the discrepancy is less than
$10^{-2}$ for most of the phases in the interval of
$[-10,70]$ days in the light curve (see figure \ref{template}).} 
From this very simple exercise we can infer how uniform the population of
Branch-normal SNe Ia {really is}{, since it} is possible to decrease
considerably the rest-frame magnitude 
standard deviation (after the maximum flux) of the {light curves in our}
sample using {just the single stretch parameter} (cf. figure \ref{sigma}).

It is worth noting that after the discovery of the correlation between SN Ia luminosities 
and the width of their light curves, other secondary empirical correlations were also 
discovered, such as the one between a SN Ia luminosity and its color \cite{Lira95,Riess96} (the brightest SNe Ia 
are also the bluest ones). The process of standardization nowadays is, therefore,
done through computational codes such as SALT2 {\cite{Guy07}} and MLCS2k2,{\cite{Jha07}} which take into account 
all these correlations.

{Our current analysis does not yet address the relationship 
between the stretch parameter and the actual {absolute} value of the peak luminosity of SNe Ia, which is a necessary step for their use  
as extragalactic distance indicators. We leave this step for a future paper, which will take into 
account the cosmological applications of what has been presented here.}

\ack{BBS would like to acknowledge financial support from the brazilian funding agency CAPES-PNPD, grant number 2940/2011.}

\appendix
\section{Basic function transformations}
\label{sec:basic_transformations}

\begin{figure}[ht]
\begin{center}
\hspace*{-1.8cm}
\includegraphics[scale=0.4]{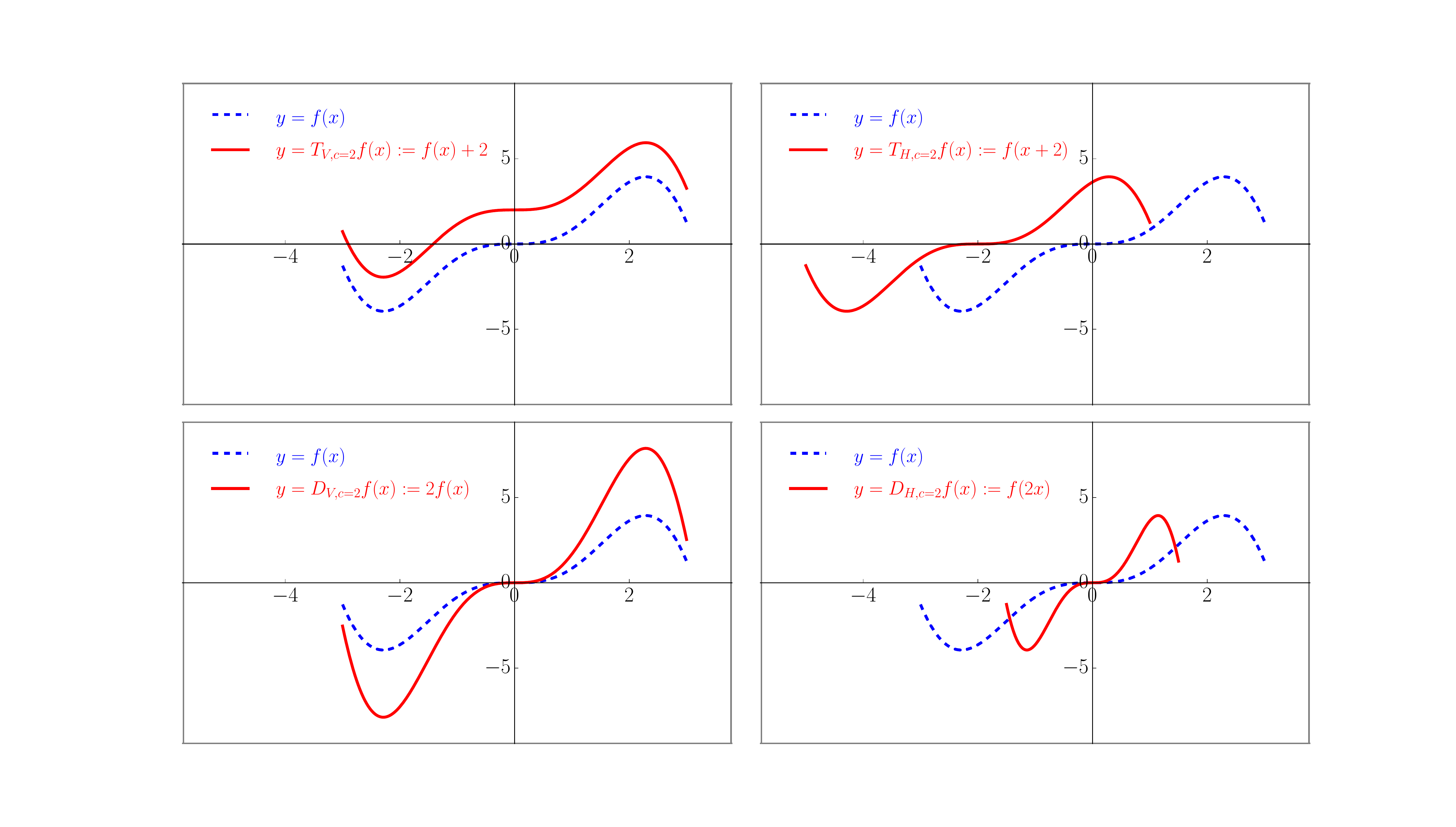}
\caption{{Basic transformations of a given arbitrary 
function $f$: the 
original graph is represented by a (blue) dashed line and the transformed graph
by a (red) solid line. Left upper panel displays the effect of a vertical
translation $T_{V,c}$ with $c=2$. Right upper panel displays the effect of a
horizontal translation $T_{H,c}$ with $c=2$. Lower left panel displays the
effect of a vertical distortion $D_{V.c}$ with $c=2$. Lower right panel displays
the effect of a horizontal distortion $D_{H.c}$ with $c=2$.}}
\label{fig:basic_function_transformations}
\end{center}
\end{figure}

{In this Appendix we investigate four simple transformations (of one real
parameter $c$ on an arbitrary function $f: x \longmapsto y=f(x)$ which bear upon the
changes on the specific flux due to distance and redshift
(cf. section~\ref{sec:dependence}). These are (cf. 
figure~\ref{fig:basic_function_transformations}):}
\begin{enumerate}
 \item {\textit{vertical translation $T_{V,c}$}:
 \begin{equation}
      T_{V,c}f: x\longmapsto y:=f(x) + c\,. \label{vert_trans}
 \end{equation}
 It always rigidly translates, along the vertical $y$-axis, the graph of the
function $f$, by $c$ ``units'': upwards, if $c>0$, and downwards, if $c<0$.}
 \item {\textit{horizontal translation $T_{H,c}$}:
 \begin{equation}
     T_{H,c}f: x\longmapsto y:=f(x+c)\,. \label{hor_trans}
 \end{equation}
 It always rigidly translates, along the horizontal $x$-axis, the graph of the
function $f$, by $c$ ``units'': left, if $c>0$, and right, if $c<0$.}
 \item {\textit{vertical distortion $D_{V,c}$}:
  \begin{equation}
     D_{V,c}f: x\longmapsto y:= cf(x)\,. \label{vert_dist}
  \end{equation}
It always distorts, along the vertical $y$-axis, the graph of the function $f$,
keeping a point with vanishing $y$ coordinate fixed: if $|c|>1$, it represents a
dilation or stretch, the more so the larger $|c|$ is, whereas if $0<|c|<1$,
it represents a contraction or compression, the more so the smaller $|c|$ is.
Furthermore, if $c<0$, this distortion is also accompanied by a reflection of
the graph with respect to the $x$-axis.}
 \item {\textit{horizontal distortion $D_{H,c}$}:
  \begin{equation}
    D_{H,c}f: x \longmapsto y:=f(cx)\,. \label{hor_dist}
  \end{equation}
It always distorts, along the horizontal $x$-axis, the graph of the function
$f$, keeping a point with vanishing $x$ coordinate fixed: if $|c|>1$, it
represents a contraction or compression, the more so the
larger $|c|$ is, whereas if $0<|c|<1$,
it represents a dilation or stretch, the more so the smaller $|c|$ is.
Furthermore, if $c<0$, this distortion is also accompanied by a reflection of
the graph with respect to the $y$-axis.}
 
\end{enumerate}

\section{Obtaining the relation between specific flux and specific luminosity}
\label{sec:basic_equation}
{Let us now proceed to the generalization of (\ref{fbolometric}) to
the Robertson-Walker spacetime, whose line element may be cast in the form:
\begin{equation}
 ds^2 = -c^2dt^2 + a(t)^2 \left[ \frac{dr^2}{1-kr^2} + r^2 \left( d\theta ^2 +
 \sin^2\theta\,d\varphi^2\right) \right]\,, \label{line_element}
\end{equation}
where $k$ is the spatial curvature and $a(t)$ is the dimensionless scale factor. 
The coordinate $r$ is variously called the comoving areal distance, transverse 
comoving distance or proper motion distance.\cite{Weinberg72,Hogg99}
}

{We first deal with the traditional case in which source and detector are
both in the Hubble flow, so that their relative velocity is all due to the
cosmic expansion, and is traditionally called a recession velocity. In this
case, time
intervals $dt_S$ in the source's rest-frame, such as the time between the
emission of 
two consecutive photons, correspond to time intervals in the detector's frame
$dt_D=(1+\bar{z})\,dt_S$, where $\bar{z}$ is the usual cosmological
redshift: $1+\bar{z}=1/a(t)$. The source-frame and detector-frame energies of
the photon will also be related by a factor $(1+\bar{z})$. Therefore, {assuming conservation of
photons}, the
bolometric flux of a source at cosmological redshift $\bar{z}$ can be
written as
\begin{equation}
f(t, r, \bar{z}, L) = \frac{L\left(\,{t}/{(1+\bar{z})}\,\right)}{4 \pi
r^2(1+\bar{z})^2}\,, \label{flux_luminosity_Hubble_flow}
\end{equation}
where  $t$ is a time coordinate measured with respect to a reference time $t_R$
which, for simplicity, we choose to be $t_R=0$ in both frames. In this way, a
time interval $t-t_R=t$ in the observer's frame corresponds to the interval
$(t-t_R)/(1+\bar{z})=t/(1+\bar{z})$ in the
source's rest-frame.}

{
Finally, we state that an equation of this same form holds for
an arbitrary motion (not in the Hubble flow) of the source and the detector,
viz.,
\begin{equation}
 f(t, r, z, L) = \frac{L\left(\,{t}/{(1+{z})}\,\right)}{4 \pi
r^2(1+{z})^2}\,, \label{flux_luminosity}
\end{equation}
where now $z$ is the total redshift between the source and the
detector, which is the really observed one\footnote{This demonstration, besides some alternative
definitions of luminosity distance, and its consequences for SNe Ia, will be
explored in a forthcoming paper, in preparation.}
Since the specific flux is related to the bolometric one by
(\ref{fbolometric}), it is obvious that it can be expressed as}
\begin{equation}
f_\lambda(\lambda, t, r, z, L_\lambda) = \frac{L_\lambda \left (
\,{\lambda}/{(1 + z)}, {t}/{(1+z)}\, \right)}{(1 + z)^3 4 \pi r^2},
\label{specific_flux}
\end{equation}
where $L_\lambda(\,{\lambda}/{(1+z)},{t}/{(1+z)}\,)$ is the specific 
luminosity in
the source's rest-frame.  

We can also obtain the frequency representation of the specific flux
as
\begin{equation}
f_\nu(\nu, t, r, z, L_\nu) = \frac{L_\nu \left (\, \nu(1+z), {t}/{(1+z)}\,
\right)}{(1+z) 4 \pi r^2}.
\label{specific_flux_nu}
\end{equation}

\noindent We can obtain (\ref{bolometric}) by integrating either
\ref{specific_flux}) on $\lambda$ or (\ref{specific_flux_nu}) on $\nu$.


\section*{References}
\begin{thebibliography}{99}

\bibitem{Hoyle60} {Hoyle F and Fowler W A 1960 Nucleosynthesis in Supernovae
\textit{Astrophys. J.} \textbf{132} 565.}

\bibitem{Padmanabhan01} Padmanabhan T 2001 Theoretical astrophysics II: stars and stellar 
systems (Cambridge: Cambridge University Press).

\bibitem{Chandrasekhar31} Chandrasekhar S 1931 The maximum mass of 
ideal white dwarfs \textit{Astrophys. J.} \textbf{74} 81--82.

\bibitem{Diemer13} Diemer B, Kessler R, Graziani C, Jordan G C IV, Lamb D Q, 
Long M and van Rossum D R 2013 Comparing the light curves
of simulated type Ia supernovae with observations using data-driven models 
\textit{Astrophys. J.} \textbf{773} 119.

\bibitem{Hillebrandt13} Hillebrandt W, Kromer M, R\"opke F K and Ruiter A J 2013 
Towards an understanding of type Ia supernovae from a synthesis of 
theory and observations \textit{Front. Phys.} \textbf{8} 116.

\bibitem{Maoz13} {Maoz D, Mannucci F and Nelemans G Observational 
clues to the progenitors of type Ia supernovae arXiv:1312.0628; to appear in 
\textit{Ann. Rev. Astron. Astrophys}.}


\bibitem{Filippenko97} Filippenko A V 1997 Optical spectra of supernovae \textit{Ann. Rev.
Astron. Astrophys.} \textbf{35} 309.

\bibitem{Branch93} Branch D, Fisher A and Nugent P 1993 On the relative
frequencies of spectroscopically normal and peculiar type IA supernovae
\textit{Astron. J.} \textbf{106} 2383.

\bibitem{Vaughan95} Vaughan T E, Branch D, Miller D L and Perlmutter S 1995
The blue and visual absolute magnitude distributions of type Ia supernovae
\textit{Astrophys. J.} \textbf{439} 558.

\bibitem{Richardson14} Richardson D, Jenkins III R L, Wright J and Maddox L 2014 
Absolute-magnitude distributions of supernovae \textit{Astron. J.} \textbf{147} 118.

\bibitem{Riess98} Riess A G \textit{et al} 1998 Observational evidence 
from supernovae for an accelerating universe and a cosmological constant
\textit{Astron. J.} \textbf{116} 1009. 

\bibitem{Perlmutter99} Perlmutter S \etal 1999 Measurements of 
$\Omega$ and $\Lambda$ from 42 high-redshift supernovae \textit{Astrophys. J.}
\textbf{517} 565.

\bibitem{Kunz12} Kunz M 2012 The phenomenological approach to modeling the 
dark energy \textit{Comptes rendus - Physique} \textbf{13} 539.

\bibitem{Dilday10} Dilday B \etal Measurements of the rate of type Ia 
supernovae at redshift z $\lesssim$ 0.3 from the SDSS-II Supernova Survey \textit{Astrophys. J.}
\textbf{713} 1026.

\bibitem{Miknaitis07} Miknaitis G \etal 2007 The ESSENCE Supernova Survey: Survey Optimization, Observations, and Supernova Photometry \textit{Astrophys. J.} \textbf{666} 674.

\bibitem{Conley11} Conley A \etal 2011 Supernova Constraints and Systematic Uncertainties from the First 3 Years of the Supernova Legacy Survey \textit{ApJS} \textbf{192} 1.

\bibitem{Sako14} Sako M \etal 2014 The Data Release of the Sloan Digital Sky Survey-II Supernova Survey arXiv:1401.3317.

\bibitem{Rest13} Rest A \etal 2013 Cosmological Constraints from Measurements of Type Ia Supernovae discovered during the first 1.5 years of the Pan-STARRS1 Survey arXiv:1310.3828.

\bibitem{Bernstein12} Bernstein J \etal 2012 Supernova Simulations and Strategies For the Dark Energy Survey \textit{Astrophys. J.} \textbf{753} 152.

\bibitem{Benitez14} Benitez N \etal 2014 J-PAS: The Javalambre-Physics of the Accelerated Universe Astrophysical Survey arXiv:1403.5237.

\bibitem{Abell09} Abell P A \etal 2009 LSST Science Book, Version 2.0 arXiv:0912.0201.

\bibitem{Carroll06}  Carroll B W and Ostlie D A 2006 \textit{An Introduction to
Modern Astrophysics} (San Fransisco: Pearson-Addison-Wesley).

\bibitem{Ostman11} \"Ostman L \etal 2011 NTT and NOT spectroscopy of
SDSS-II supernovae \textit{Astron. Astrophys.} \textbf{526} A28.

\bibitem{Bessell05} Bessell S 2005 Standard photometric systems \textit{Ann. Rev.
Astron. Astrophys.} \textbf{43} 293.

\bibitem{Oke65} Oke J B 1965 Absolute spectral energy distributions in stars 
\textit{Ann. Rev. Astron. Astrophys.} \textbf{3} 23--46.

\bibitem{Oke83} Oke J B and Gunn J E 1983 Secondary standard stars for
absolute spectrophotometry \textit{Astrophys. J.} \textbf{266} 713.

\bibitem{Contreras10} Contreras C \etal 2010 The Carnegie Supernova
Project: first photometry data release of low-redshift type Ia supernovae
\textit{Astron. J.} \textbf{139} 519--539.

\bibitem{Nugent_templates} Peter Nugent templates,
\url{<http://supernova.lbl.gov/~nugent/nugent_templates.html>}.

\bibitem{Synge60} Synge J L 1960 \textit{Relativity: the general theory} (Amsterdan: North-Holland,).

\bibitem{Narlikar94} Narlikar J V 1994 Spectral shifts in general relativity \textit{Am. J. Phys.} \textbf{62} 903.

\bibitem{Patat96} Patat F, Benetti S, Cappellaro E, Danziger I J, 
della Valle M, Mazzali P A and Turatto M 1996 The type IA supernova 1994D in NGC
4526: the early phases \textit{Mon. Not. Royal Astron. Soc.} \textbf{278} 111.

\bibitem{Branch03} Branch D, Garnavich P, Matheson T, Baron E, 
Thomas R C, Hatano K, Challis P, Jha S and Kirshner R P 2003 Optical spectra of the
type Ia supernova 1998aq \textit{Astron. J.} \textbf{126} 1489.

\bibitem{Gerardy05} Gerardy C L 2005 \textit{1604-2004:
Supernovae as cosmological lighthouses} ASP Conference Series \textbf{342} edited by 
Turatto M, Benetti S, Zampieri L and Shea W (San Franciso: Astronomical Society 
of the Pacific) p~250.

\bibitem{suspect} SUSPECT Web Site, \url{<http://nhn.nhn.ou.edu/~suspect/>}.

\bibitem{Phillips93} Phillips M M 1993 The absolute magnitudes of type Ia 
supernovae \textit{Astrophys. J.} \textbf{413} L105.

\bibitem{Hamuy96} Hamuy M, Phillips M M, Schommer R A, Suntzeff N B, Maza J and
Avil\'es R 1996 The absolute luminosities of the 
Cal\'an/Tololo type Ia supernovae \textit{Astron. J.} \textbf{112} 2391.

\bibitem{CSPWebsite} Carnegie Supernova Project Web Site, \url{<http://csp.obs.carnegiescience.edu/data>}.

\bibitem{Stritzinger11} {Stritzinger M D \etal 2011 The Carnegie
Supernova Project: 
second photometry data release of low-redshift type Ia supernovae  
\textit{Astron. J.} \textbf{142} 156.}

\bibitem{Goldhaber01} Goldhaber G \etal 2001 Timescale stretch
parameterization of type Ia supernova $B$-band light curves \textit{Astrophys.
J.} \textbf{558} 359.

\bibitem{Hayden10} {Hayden B T \etal 2010 The rise and fall of 
type Ia supernova light curves in the SDSS-II Supernova Survey 
\textit{Astrophys. J.} \textbf{712} 350--366.}

\bibitem{Lira95} Lira P, Masters thesis, University of Chile.

\bibitem{Riess96} Riess A G, Press W H and Kirshner R P 1996 A Precise Distance 
Indicator: Type IA Supernova Multicolor Light-Curve Shapes \textit{Astrophys. J.} \textbf{473} 
88.

\bibitem{Guy07} Guy J \etal 2007 SALT2: using distant supernovae to improve 
the use of Type Ia supernovae as distance indicators \textit{Astron.
Astrophys.} {\bf 466} 11--21.

\bibitem{Jha07} Jha S, Riess A G and Kirshner R P 2007 Improved Distances to Type 
Ia Supernovae with Multicolor Light Curve Shapes: MLCS2k2
\textit{Astrophys. J.} {\bf 659} 122.

\bibitem{Weinberg72} Weinberg S 1972 \textit{Gravitation and cosmology: principles
and applications of the general theory of relativity} (New York: John Wiley \& Sons).

\bibitem{Hogg99} Hogg D W 1999 Distance measures in
cosmology arXiv:astro-ph/9905116.

\end{thebibliography}
\end{document}